\documentclass[12pt,preprint]{aastex}
\usepackage{graphicx}
\usepackage{amssymb}

\begin{document}
\title{Starburst-Driven Galactic Winds: Filament Formation and Emission Processes}

\author{Jackie L. Cooper, Geoffrey  V. Bicknell, and Ralph  S. Sutherland}
\affil{Research School of Astronomy and Astrophysics, \\ 
  Australian National University, ACT 0200, Australia}
\email{jcooper@mso.anu.edu.au}
 \and
\author{Joss Bland-Hawthorn}
\affil{Institute of Astronomy, School of Physics, \\
University of Sydney, NSW 2006, Australia}

\begin{abstract}

We have performed a series of three-dimensional simulations of the interaction of 
a supersonic wind with a non-spherical radiative cloud. These simulations are 
motivated by our recent three-dimensional model of a starburst-driven  galactic 
wind interacting with an inhomogeneous disk, which show that an optically
emitting filament can be formed by the break-up and acceleration of a cloud
into a supersonic wind. In this study we consider the evolution of a cloud
with two different geometries (fractal and spherical) and investigate the
importance of radiative cooling on the cloud's survival. We have also
undertaken a comprehensive resolution study in order to ascertain the effect 
of the assumed numerical resolution on the results. We find that the ability of 
the cloud to radiate heat is crucial for its survival, with a radiative cloud 
experiencing a lower degree of acceleration and having a higher relative Mach number 
to the flow than in the adiabatic case. This diminishes the destructive effect
of the Kelvin-Helmholtz instability on the cloud. While an adiabatic cloud is
destroyed over a short period of time, a radiative cloud is broken up via the
Kelvin-Helmholtz instability into numerous small, dense cloudlets, which are
drawn into the flow to form a filamentary structure. The degree of
fragmentation is highly dependent on the resolution of the simulation, with
the number of cloudlets formed increasing as the Kelvin-Helmholtz instability
is better resolved. Nevertheless, there is a clear qualitative trend, with the
filamentary structure still persistent at high resolution. The geometry of the
cloud effects the speed at which the cloud fragments; a wind more rapidly
breaks-up the cloud in regions of least  density. A cloud with a more
inhomogeneous density distribution fragments faster than a cloud 
with a more uniform structure (e.g. a sphere). We confirm the mechanism behind
the formation of the H$\alpha$ emitting filaments found in our global
simulations of a starburst-driven wind. Based on our resolution study, we
conclude that bow shocks around accelerated gas clouds, and their interaction,
are the main source of the soft X-ray emission observed in these
galactic-scale winds. 
 
\end{abstract}

\section{INTRODUCTION}

The interstellar medium is known to be inhomogeneous, consisting of various
gaseous phases, from cool molecular clouds to tenuous million degree gas
\citep{Cox2005}. A study of the interaction of a supersonic wind with a
dense cloud is a problem with many astronomical applications, such as galactic
winds, supernova remnants, and broad absorption line quasars, and has received much 
attention. In the past this interaction has been studied both analytically
\citep[e.g.][]{Mckee1975,Heathcote1983} and numerically
\citep[e.g.][]{Sgro1975,Woodward1976,Nittmann1982}. Over the last two decades,
numerous two- and three- dimensional simulations have been performed. Many
early attempts assumed an adiabatic interaction 
\citep[e.g.][]{Stone1992,Klein1994,Xu1995}, 
while later attempts have included radiative cooling
\citep[e.g.][]{Mellema2002,Melioli2004,Fragile2004,Fragile2005,Marcolini2005,Melioli2005,Tenorio-Tagle2006,Orlando2005,Orlando2006,Orlando2008}. Among the plethora of
simulations reported in the literature, the effects of thermal conduction
\citep[e.g.][]{Marcolini2005,Orlando2005,Orlando2006,Orlando2008,Recchi2007},
magnetic fields
\citep[e.g.][]{MacLow1994,Gregori1999,Gregori2000,Fragile2005,Orlando2008,Shin2008},
photoevaporation \citep[e.g][]{Melioli2005,Raga2005,Tenorio-Tagle2006}, and
the presence of multiple clouds
\citep[e.g.][]{Poludnenko2002,Melioli2005,Pittard2005,Tenorio-Tagle2006} have been
considered. The wind/cloud interaction has also been investigated in the laboratory via laser experiments
\citep[e.g.][]{Klein2003,Hansen2007}. In this paper, we report on our high resolution three-dimensional 
simulations of the interaction of a supersonic wind with a non-uniform radiative 
cloud. This work is motivated by our recent simulations of a starburst-driven 
galactic wind \citep[][;~hereafter paper {\sc i}]{Cooper2008}, which showed
that a optically emitting filament could be formed by the break-up and
ram-pressure acceleration of a cloud into a supersonic wind. These simulations 
were three-dimensional, radiative, and incorporated an inhomogeneous disk that 
allowed us to study the interaction of a galactic scale wind with fractal clouds of 
disk gas. 

The interaction of a spherical cloud with a shock wave is often characterized
by four evolutionary phases \citep[e.g.][; and references therein]{Klein1994}:
\begin{enumerate}
\item An initial phase where the blast wave interacts with the cloud. A shock
  passes into the cloud, with another shock reflected into the surrounding medium.
\item A phase of shock compression where the flow around the cloud converges
  on the axis at the rear. During this phase, the cloud begins to
  flatten, with its transverse size greatly reduced. A shock is  also driven into
  the back of the cloud.
\item The re-expansion phase where the shock transmitted into the cloud has 
reached the back surface and produces a strong rarefaction back into the cloud. 
This leads to expansion of the shocked cloud downstream.
\item A final phase where the cloud fragments and is destroyed by 
hydrodynamical instabilities (e.g. Kelvin-Helmholtz and Rayleigh-Taylor 
instabilities). 
\end{enumerate}

\citet{Klein1994} performed an extensive series of two-dimensional simulations
of an spherical adiabatic cloud interacting with a shock wave. They found that
irrespective of the initial parameters, the cloud was destroyed within several
cloud crushing times. \citet{Xu1995} confirm this result
in their three-dimensional simulations. However, \cite{Mellema2002} in a short 
letter, reported their two-dimensional study of the evolution a cloud in a radio 
galaxy cocoon. Their simulations included the effects of radiative cooling and 
showed that merging of the front and back shocks leads to the formation of an 
elongated structure. This structure breaks-up into fragments which are not 
immediately destroyed. This increased ability for the cloud to survive has been 
reproduced by all studies that implement radiative cooling in their simulations. 
In addition, it has been shown that strong cooling in the cloud can cause a thin 
dense shell that acts to protect the cloud from ablation 
\citep{Fragile2004,Melioli2005,Sutherland2007}. 

Thermal conduction can suppress the hydrodynamical instabilities that act to
fragment a cloud \citep{Orlando2005,Marcolini2005}, while the presence of a 
magnetic field has been shown to both hasten \citep[e.g][]{Gregori1999} and delay 
\citep[e.g.][]{MacLow1994,Fragile2005} the cloud's destruction. More recently, 
\citet{Orlando2008} performed three-dimensional simulations of the wind/cloud 
interaction that included the effects of radiative cooling, thermal conduction, 
and magnetic fields. They showed that in the presence of an ambient magnetic 
field, the effect of thermal conduction in stabilizing the cloud may be diminished 
depending on the alignment of the field. Clearly the survival of a cloud 
interacting with a strong wind is a problem of significantly more complexity than 
indicated by the purely adiabatic scenario investigated by \citet{Klein1994}
and others. In order to compare the results of our current study to those of our
global model we exclude the thermal conduction, magnetic fields, or
photoevaporation. However, we consider the possible effect of these phenomena
on our results in \S~\ref{missing}. 

In this work, we further investigate the effects of radiative cooling on the
survival of the  cloud, by performing high $\sim 0.1$ pc resolution
simulations of both a radiative and an adiabatic cloud, allowing us to perform
a direct comparison between the two scenarios. Symmetry is {\em not} assumed,
with the entire cloud being modeled, in contrast to the strategy employed in
many previous three-dimensional models. This allows us to fully investigate of
the turbulent flow on the evolution of the cloud. In order to understand the
effects of the cloud's structure on the wind/cloud interaction, we consider
two different cloud geometries: a more realistic fractal cloud, similar to
those that comprised the inhomogeneous disk in paper {\sc i}, and the
idealized case of a spherical cloud. The effect of the cloud's geometry on its
evolution has been investigated before. \citet{Xu1995} considered the
interaction of a shock wave with a spherical and two different prolate cloud
geometries, each with a different alignment of the cloud's major axis. In addition,
\citet{Mellema2002} considered  both spherical and elliptical cloud
geometries. Both studies show that the initial geometry of the cloud can alter
the evolution of the cloud significantly.
  
One of the most important results from paper {\sc i} was our suggestion of a
mechanism for the formation of the filaments seen in starburst-driven winds at
optical wavelengths \citep[see][;~for review]{Veilleux2005}. According to
paper {\sc i} the filaments are formed from clouds of disk gas that have been
accelerated into the outflow by the ram-pressure of the wind. An important
question that arises is: Can the clouds survive being immersed in a hot
supersonic wind long enough to form a filament and remain sufficiently cool to
emit at optical temperatures? Or will they be heated and destroyed by
hydrodynamical instabilities? Here we set out to answer this question.   

Another significant result to arise from paper {\sc i} is our proposal of 
four different mechanisms that would give rise to soft X-ray emission that is 
spatially correlated with the filamentary optical emission; a major finding of 
recent {\it Chandra} observations 
\citep{Strickland2002,Strickland2004a,Strickland2004b,Cecil2002,Martin2002}. Our 
global simulations found that soft X-rays can arise from (i) mass-loading from 
ablated clouds, (ii) the intermediate temperature interface between the hot wind 
and cool filaments, (iii) bow shocks upstream of clouds accelerated into the 
outflow, and (iv) interactions between these bow shocks. The first two mechanisms 
involve the mixing of hot and cold gas, and are possibly caused by numerical 
diffusion in the simulations, and thus may not be physical. To investigate this 
possibility, we have performed a detailed resolution study of the wind/cloud 
interaction. The study also allows us to test the impact of the numerical 
resolution on the evolution of the cloud. 

\section{NUMERICAL METHOD}\label{method}

\subsection{Description of the Code}

The simulations were performed using the PPMLR (Piecewise Parabolic Method
with a Lagrangian Remap) code utilized in paper {\sc i}. PPMLR is a 
multidimensional hydrodynamics code based on the method described by 
\citet{Colella1984} and has been extensively modified 
\citep[e.g.][]{Sutherland2003a,Sutherland2003b} from the original VH-1 code 
\citep{Blondin1995}. Thermal cooling, based on output from the MAPPINGS III code 
\citep{Sutherland1993} has been implemented. This allows for a realistic evolution 
of a radiatively cooling gas \citep{Sutherland2003b,Saxton2005}. The simulations 
discussed in this paper are three-dimensional and utilize Cartesian (x,y,z)
coordinates. They were performed on the SGI Altix computer operated by the the 
Australian Partnership for Advanced Computing.

\subsection{Problem Setup}

In order to study the wind/cloud interaction in sufficient detail, whilst still
retaining the ability to follow the evolution and survival of the cloud over a 
period of approximately 1 million years, the simulations cover a physical spatial 
range of 50 $\times$ 50 $\times$ 150 pc, with the cloud centered on the origin. 
As our intent is to compare these simulations to the formation and survival of the 
clouds found in paper {\sc i}, we choose initial conditions  for the
wind based upon the results of those simulations (Table \ref{init_cond}). In order 
to understand our choice, we briefly recount the formation and evolution of the 
starburst-driven winds which we simulated in paper {\sc i}:
\begin{enumerate}
\item A series of small bubbles of hot ($T \gtrsim 10^7 K$) gas form in the
      starburst region. As these bubbles expand, they merge and follow the path of
      least resistance out of the disk of the galaxy, i.e. the tenuous hot gas
      surrounding the denser disk clouds.
\item As the bubble breaks out of the disk, it begins sweeping up the the
      surrounding halo gas entering the ``snow-plow'' phase of its evolution.
      The structure of the wind in the phase is                  
      characterized by 5 different zones: (i) the injection zone, (ii) a 
      supersonic free wind, (iii) a region of hot, shocked turbulent gas, (iv) a 
      cooler ``shell'' of swept-up halo gas, and (v) the undisturbed ambient gas.
\item Clouds of disk gas inside and surrounding the central injection zone are
      broken-up by the freely expanding wind; fragments of disk gas are
      accelerated into the outflow by the ram-pressure of the wind.
\end{enumerate}
Since it is the interaction of the clouds with the freely expanding wind that
results in their fragmentation, we select the temperature, density and
velocity of our supersonic wind to be that of the inner free-wind region,
namely $T_{\rm w} = 5 \times 10^{6} \rm~K$, $n_{\rm w} = 0.1 \rm~cm^{-3}$, and
$v_{\rm w} = 1200 \rm~km~s^{-1}$ respectively.

\begin{deluxetable}{lcc}
\tablewidth{0pt}
\tablecaption{Initial Conditions \label{init_cond}}
\tablehead{
\colhead{Parameter} & \colhead{Symbol}  & \colhead{Value}}
\startdata
Wind Temperature (K) & $T_{\rm w}$ & $5 \times 10^{6}$ \\
Wind Density ($\rm cm^{-3}$) & $n_{\rm w}$ & 0.1 \\
Wind Velocity ($\rm km~s^{-1}$) & $v_{\rm w}$ & $1200$ \\
Wind Mach Number & $\mathcal{M}_{\rm w}$ & 4.6 \\
Average Cloud Temperature (K) & $T_{\rm c}$ & $5 \times 10^{3}$ \\
Cloud Velocity ($\rm km~s^{-1}$) & $v_{\rm c}$ & 0 \\
Cloud Radius (pc) & $r_{\rm c}$ & 5 \\
Fractal Cloud Volume ($\rm pc^{3}$) & $V_{\rm c\_frc}$ & 1491 \\
Spherical Cloud Volume ($\rm pc^{3}$) & $V_{\rm c\_sph}$ & 523 \\
\enddata
\end{deluxetable}

Our simulations make use of two cloud geometries: a fractal shaped
and a spherical shaped cloud. The fractal cloud was chosen in order to allow us to 
compare the break-up of the clouds in this study to the results of our global 
model in paper {\sc i} and has the same form as the clouds in the 
inhomogeneous disk used in the global study. The use of the fractal cloud also 
allows us to investigate the effects of inhomogeneities in the clouds's density 
distribution on its evolution. A spherical cloud was also modeled in order to 
allow us to better understand the importance of the assumed cloud geometry. To 
create the fractal cloud we first created a 1024 $\times$ 1024 $\times$ 1024 sized 
fractal cube using the method described in \citet{Sutherland2007} and paper {\sc 
i}. A single cloud was isolated and extracted from this cube using a 
blob-coloring technique where each cell is examined, and any discontinuous group 
of non-zero cells given a unique label. An appropriate cloud was then selected and
placed at cell number (256,256,256) of a 512 $\times$ 512 $\times$ 1536 sized 
grid, i.e. the origin of the simulation. The smaller resolution grids used in our 
fractal simulations were created by downsizing this larger grid (see Table 
\ref{sim_param}). These arrays represent the density of the cloud. 

The initial grid was setup by first setting the density ($n_{\rm w}$) and pressure 
($P_{\rm w} = n_{w}kT_{w}/{\mu}m_{p}$) to be that of the hot wind. The fractal 
cloud was then created by adding the density array representing the cloud to the 
density of the halo gas. To create the spherical cloud, a high density spherical
region of radius $r_{\rm c} = 5 \rm~pc$ was centered on the origin of the
computational grid. The radial profile of the density of the spherical cloud is 
described by an exponential with a scaling radius of 3, in order to mimic the 
tapered density distribution in the fractal cloud's core. A similar tapered density
distribution was considered by \citet{Nakamura2006} for an adiabatic
cloud. The boundary condition on the inner z axis was set to have a fixed
inflow with the same properties as that of the hot wind (e.g. $T_{\rm w} = 5
\times 10^{6} \rm~K$, $n_{\rm w} = 0.1 \rm~cm^{-3}$, and $v_{\rm w} = 1200
\rm~ km~s^{-1}$). All other boundaries were set to be inflowing/outflowing. 

\begin{figure}[tbp]
\begin{center}
\epsscale{0.75}
\plottwo{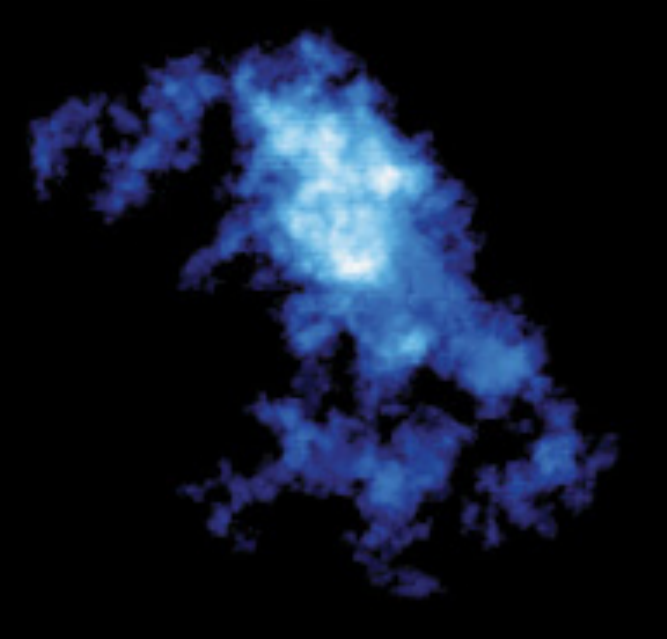}{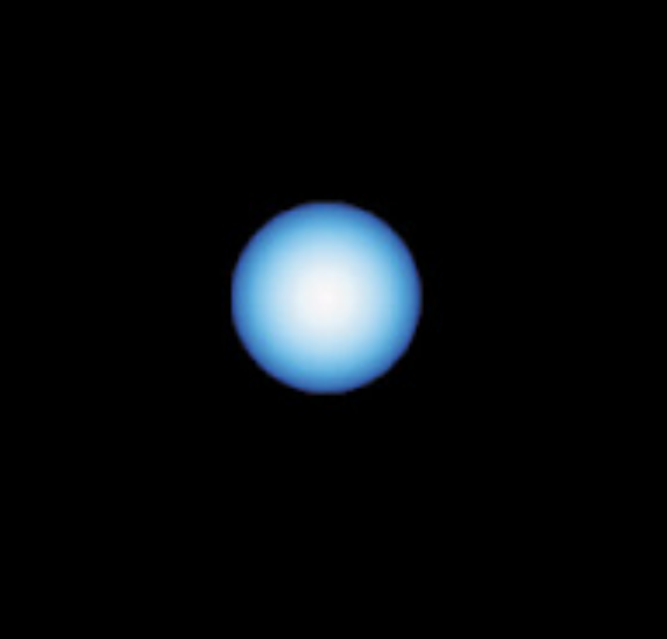}
\caption{Volume renderings of the projected density showing the initial
  distribution of the fractal (left) and spherical (right) clouds.}\label{fig:dens_profile}
\end{center}
\end{figure}

Figure \ref{fig:dens_profile} shows the initial density distribution
of both the fractal and spherical clouds. The average number
density of the fractal cloud is set to be $n_{\rm c} = 63 \rm ~cm^{-3}$ and
has a total mass of $M_{\rm c} = 1387 \rm ~M_{\odot}$, occupying a volume of
$V_{\rm c\_frc} = 1491 \rm ~pc^{3}$. The spherical cloud is setup to occupy a
similar volume to the dense core of the fractal cloud, having a radius of
$r_{\rm c} = 5 \rm~pc$ and occupying a volume of $V_{\rm c\_sph} = 523 \rm
~pc^{3}$. The average density of the spherical cloud is $n_{\rm c} = 91 \rm
~cm^{-3}$ and has a total mass of $M_{\rm c} = 523 \rm ~M_{\odot}$. The lower
average density of the fractal cloud is due to the large volume of less dense
($n_{\rm c} = 30 \rm ~cm^{-3}$) gas that surrounds the cloud core (Fig.
\ref{fig:dens_profile}; left panel). In order to understand the effect of the
cloud's assumed initial density on its evolution, a simulation in which the
density of the fractal cloud was doubled ($n_{\rm c} = 126 \rm ~cm^{-3}$) was
also performed. In all simulations, the temperature and velocity of each cloud was 
set to be $T_{\rm c} = 5 \times 10^{3} \rm~K$ and $v_{\rm c} = 0 \rm ~km~s^{-1}$
respectively.

While these simulations were designed to be applicable to starburst-driven winds 
and therefore have densities and temperatures typically found in a such an 
environment, they are also applicable to other astronomical phenomena that involve 
the interaction of a supersonic wind with a cloud of gas (e.g supernova remnants). 
When cooling is included, the PPMLR code has a one parameter scaling, which is 
discussed in \citet{Sutherland2007}. In general, the density scale of the 
simulations is inversely proportional to the spatial scale, so that within reason, 
these simulations can be adapted to problems on both larger and small scales. 

\subsection{The Simulations}

The interaction of a supersonic wind with a cloud of dense gas is a problem of some
complexity. Whilst there are many factors that could affect a clouds
evolution and survival, such as thermal conduction 
\citep{Marcolini2005,Orlando2005}, magnetic fields \citep{Fragile2005} and 
photoevaporation \citep{Tenorio-Tagle2006} (see \S~\ref{missing}), here
we focus on the importance of 
radiative cooling and the effect on the clouds initial structure. We have also 
performed a comprehensive resolution study in order to ascertain the degree as to 
which the assumed resolution effected our global simulations in paper {\sc i}. 

We adopt the following naming convention for our simulations: An $r$ or an $a$ 
indicates whether the simulation includes radiative cooling or is adiabatic, while 
an $f$ or an $s$ indicates if the geometry of the cloud is fractal or spherical 
respectively. The numerical value indicates the number of cells in the x-plane of 
the computational grid. For example, the simulation rf384 includes radiative 
cooling, incorporates a fractal cloud and utilizes a computational  grid of size 
$384 \times 384 \times 1152$ cells. An exception to the  naming convention is 
rfd384, which is identical to rf384, but whose cloud is twice as dense. 

\begin{deluxetable}{ccccccc}
\tablewidth{0pt}
\tablecaption{Simulation Parameters \label{sim_param}}
\tablehead{
\colhead{Model} & \colhead{Grid Size}  & \colhead{Resolution
  (pc)} & \colhead{$n_{\rm c}\tablenotemark{a} ~\rm ~(cm^{-3})$} &
  \colhead{$M_{\rm c}\tablenotemark{b} ~\rm ~(M_{\odot})$} & 
  \colhead{Cooling?} & \colhead{Shape\tablenotemark{c}}}
\startdata
rf064 & $64 \times 64 \times 192$ & 0.78 & 63 & 1387 & yes & F \\
rf096 & $96 \times 96 \times 288$ & 0.52 & 63 & 1387 & yes & F \\
rf128 & $128 \times 128 \times 384$ & 0.39 & 63 & 1387 & yes & F \\
rf192 & $192 \times 192 \times 576$ & 0.26 & 63 & 1387 & yes & F \\
rf256 & $256 \times 256 \times 768$ & 0.20 & 63 & 1387 & yes & F \\
rf384 & $384 \times 384 \times 1152$ & 0.13 & 63 & 1387 & yes & F \\
rf512 & $512 \times 512 \times 1536$ & 0.10 & 63 & 1387 & yes & F \\
af384 & $384 \times 384 \times 1152$ & 0.13 & 63 & 1387 & no & F \\
rs384 & $384 \times 384 \times 1152$ & 0.13 & 91 & 703 & yes & S \\
as384 & $384 \times 384 \times 1152$ & 0.13 & 91 & 703 & no & S \\
rfd384 & $384 \times 384 \times 1152$ & 0.13 & 126 & 2770  & yes & F \\
\enddata
\tablenotetext{a}{Average cloud density}
\tablenotetext{b}{Cloud mass}
\tablenotetext{c}{Cloud Shape: F = fractal, S = spherical}
\end{deluxetable}

In total, eleven simulations were performed, with the
purpose  of each falling within the four different categories outlined below.   
\begin{enumerate}
\item {\it Resolution Study} -  Models {\bf rf064}, {\bf rf096}, {\bf
  rf128}, {\bf rf192}, {\bf rf256}, {\bf rf384}, and {\bf rf512} form our
  resolution study. These seven simulations all include radiative cooling and  
  utilize a fractal cloud. The resolution of each simulation is given in Table 
  \ref{sim_param} and ranges from 0.79 to 0.10 pc per cell width.   
\item {\it Cloud Structure} - Model {\bf rs384} has a resolution of 0.13 pc per
  cell width, includes radiative cooling and utilizes a spherical cloud. By
  comparison to rf384, this model is designed to investigate the effect of
  the shape of the cloud on its break-up and survival.
\item {\it Radiative Cooling} - Models {\bf af384} and {\bf as384} have the
  same properties as rf384 and rs384 respectively, but are adiabatic in nature. 
  Both models are designed to help understand the importance of radiative cooling 
  on the evolution and survival of a cloud.
\item {\it Cloud Density} - Model {\bf rfd384} is identical to rf384, but has a
  larger cloud density and mass of $n_{\rm c} = 126 \rm ~cm^{-3}$ and $M_{\rm c} = 
  2770 \rm ~M_{\odot}$ respectively. This model is designed to investigate the 
  effect of the clouds initial density on its evolution.
\end{enumerate}
A summary of the parameters used in each simulation is given in Table 
\ref{sim_param}.

For each simulation we record the density, temperature, pressure, velocity,
emissivity, and a cloud gas tracer in each cell at intervals of 0.01 Myr.   
With the exception of rf512, each simulation is followed until the time at
which the cloud flows off computational grid. In the case of rf512, a
computational error occurred while the simulation was in progress. Unfortunately 
this simulation is too computationally expensive to re-run, and we are therefore 
only able to follow the evolution of rf512 to a time t = 0.7 Myr. As such, our 
resolution study only considers the first 0.7 Myr of the evolution.

Based on their two-dimensional adiabatic simulations, \citet{Klein1994} suggest
that a resolution of 120 cells per cloud radius is necessary in order to
sufficiently capture the hydrodynamics of the wind/cloud interaction. While
this proposed benchmark is easily achieved for a two-dimensional study, it
becomes computationally demanding for a fully three-dimensional simulation that
include more complicated physics (e.g. radiative cooling, thermal
conduction). In order to overcome this difficulty, other authors have
performed two-dimensional axisymmetric simulations \citep{Orlando2008},
assumed symmetry in the solution
\citep{Gregori1999,Gregori2000,Melioli2005,Orlando2005}, and 
neglected the more complex physics in their three-dimensional calculations
\citep{Orlando2005}. We {\em do not} assume symmetry as we show in
\S~\ref{evolution} that, even in the idealized case of a spherical cloud, the
solution becomes asymmetric as the cloud is broken-up and accelerated into the
turbulent flow. As the PPMLR code used in this work and paper {\sc i} utilizes
a uniform grid, the computational resources required to model the entire cloud
at the 120 cells per cloud radius resolution suggested by \citet{Klein1994}
would be excessive. We are forced to limit the main simulations in this study
to a resolution of 38 cells per cloud radius (we consider a resolution of 50
cells per cloud radius in our resolution study). Nevertheless, it has been
shown that the global properties of the interaction, as well as the averaged
characteristics of cloud ablation process, can be well described at
resolutions below this criterion \citep[see, for
  example,][]{Gregori2000,Poludnenko2002,Melioli2005}.  

\section{EVOLUTION}\label{evolution}

\subsection{Description of the Wind/Cloud Interaction}

\subsubsection{Spherical Cloud}

\begin{figure}[tbp]
\epsscale{0.8}
\plotone{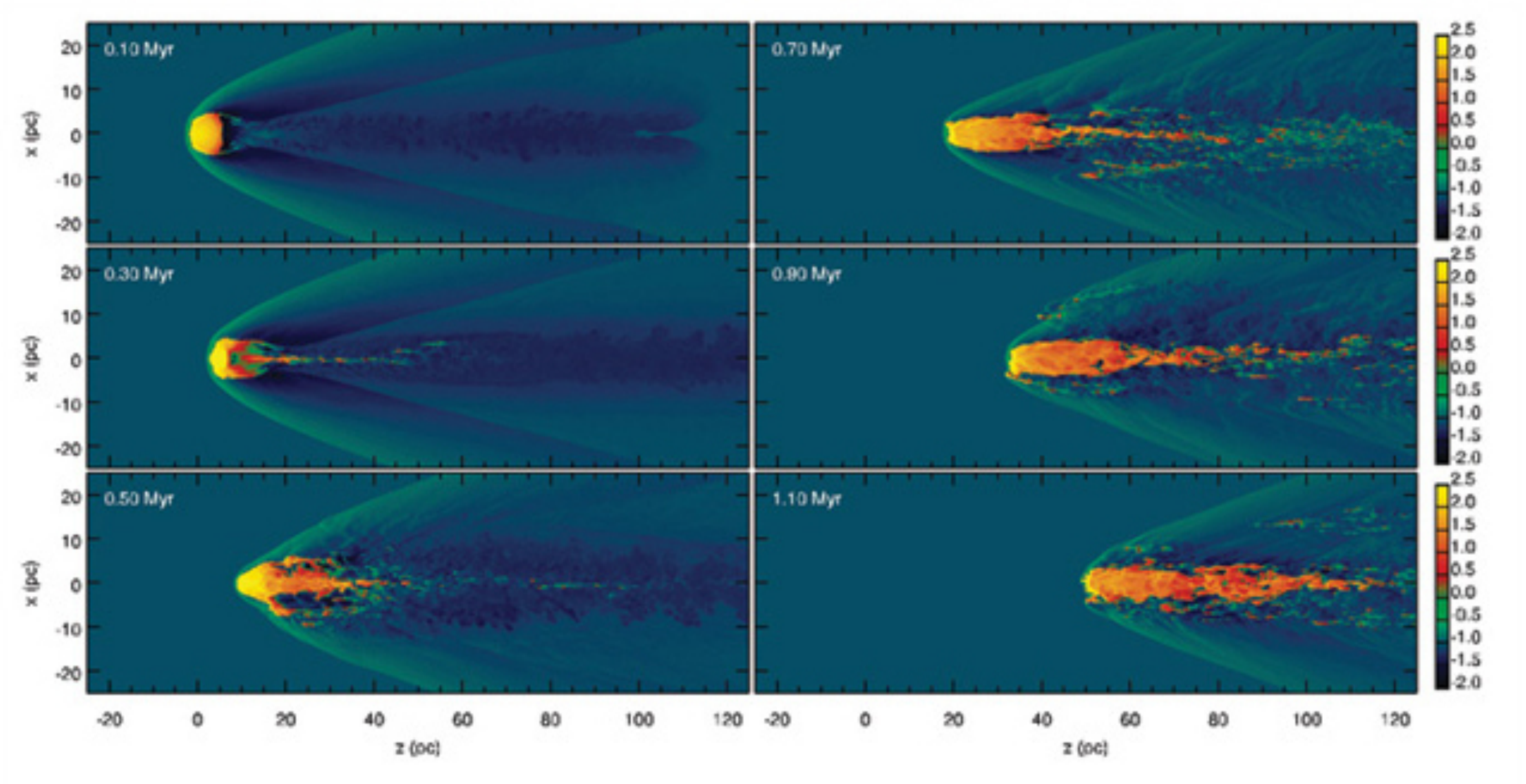}
\caption{Logarithm of the density through the y=0 plane in model rs384 showing the
  evolution of a radiative spherical cloud.}\label{fig:dens_sph}
\end{figure}

\begin{figure}[tbp]
\epsscale{0.8}
\plotone{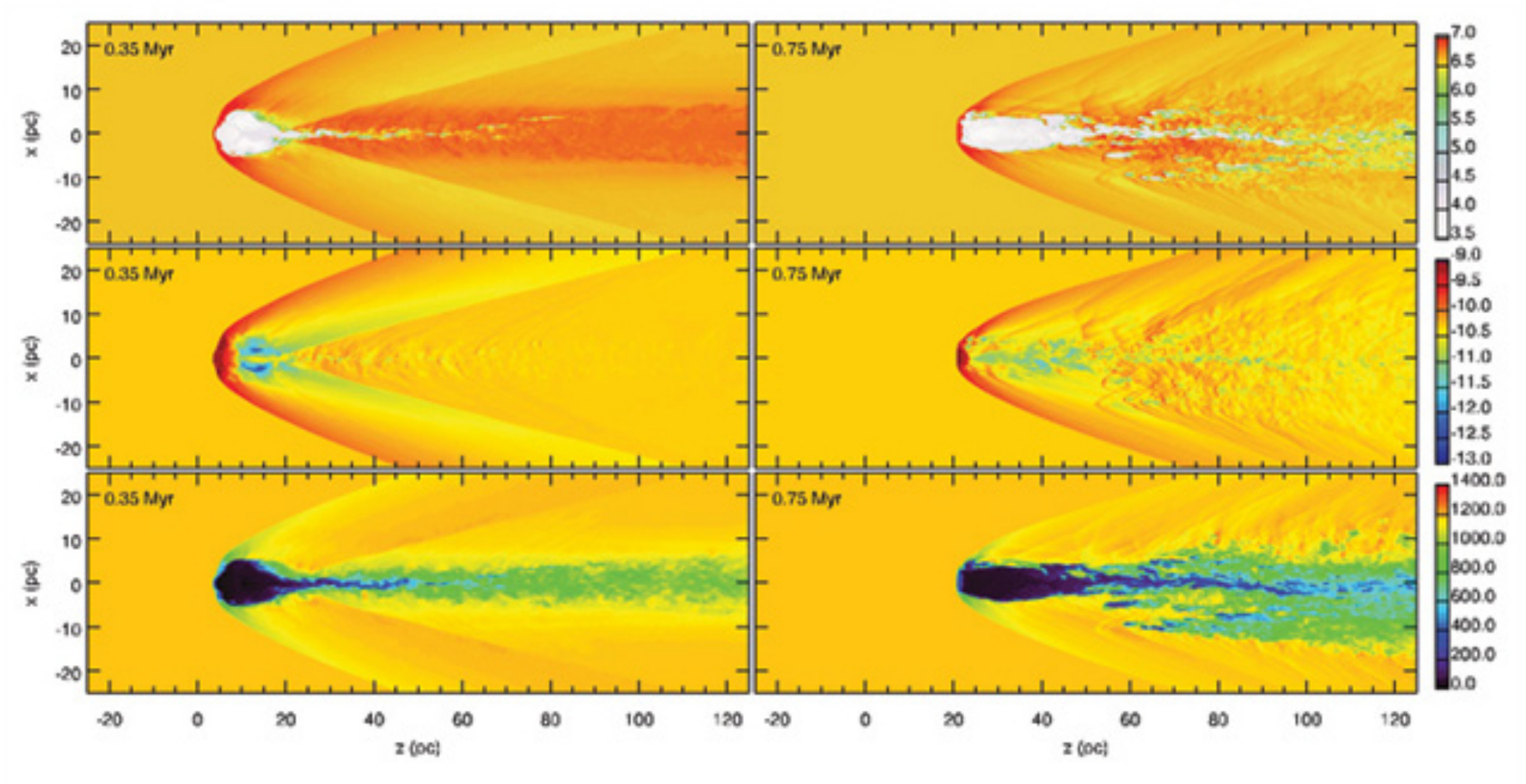}
\caption{Log-temperature (top), log-pressure (middle), and velocity (bottom)
  at 0.35 (right) and 0.75 (left) epochs through the y=0 plane in model
  rs384.}\label{fig:var_sph}
\end{figure}

We start with the simple case of the interaction of a spherical radiative cloud 
(rs384) with a supersonic wind, before describing the evolution of the fractal 
cloud. In order to illustrate the hydrodynamics of the wind/cloud interaction, 
throughout this section we will use slices of the density, temperature, pressure 
and velocity through the central y=0 plane. Figure \ref{fig:dens_sph} shows the 
evolution of the spherical cloud a 6 different epochs from 0.1 to 1.1 Myr at 0.2 
million year intervals. Each panel represents the logarithm of the number density 
($\rm cm^{-3}$). The top panels of Figure \ref{fig:var_sph} show the logarithm of 
the temperature ($\rm K$), the middle panels show the logarithm of the pressure 
($\rm cm~s^{-2}$), and the bottom panels show the magnitude of the velocity ($\rm
km~s^{-1}$). The evolution at two different epochs is given: 0.35 Myr (left 
panels) and 0.75 Myr (right panels).  

The spherical cloud is initially at rest and is immersed in a hot ($T = 10^6 ~\rm 
K$), supersonic ($V = 1200 ~\rm km~s^{-1}$) wind. A bow shock is immediately 
formed upstream of the cloud. At 0.10 Myr (Fig. \ref{fig:dens_sph}; upper left 
panel), the ``front'' of the cloud has been exposed to the high pressure of the 
shock, while gas is ablated from the ``back'' of the cloud a result of the 
Kelvin-Helmholtz instability and the strong rarefaction that is formed. A high
density shock ($n = 1000 ~\rm cm^{-3}$) begins to propagate through the cloud, 
reflecting of the back wall at approximately 0.30 Myr (Fig. \ref{fig:dens_sph}; 
middle left panel). The Kelvin-Helmholtz instability continues to work, stripping 
material from the edge of the cloud (e.g. 0.5 - 1.1 Myr). This material is
funneled to approximately 5 pc behind the cloud where it combines and
condenses forming a tail of dense ($n \sim 10 ~\rm cm^{-3}$), cool ($T ~\sim
10^4 ~\rm K$) cloudlets, which are entrained into the hot turbulent flow
downstream from the main cloud (Fig. \ref{fig:var_sph}; upper left panel).   

As the cloud evolves (Figs. \ref{fig:dens_sph}; right 
panels), the tail of cool gas continues to grow, becoming thicker as the cloud 
elongates. The cloud material appears as a filament of cool, low velocity ($v < 
400 ~\rm km~s^{-1}$) gas, immersed inside a region of hot ($T > 10^7 ~\rm K$), 
turbulent gas with velocity $v \sim 400 - 1000 ~\rm km~s^{-1}$. Small cloudlets 
continue to be broken off the main cloud as the Kelvin-Helmholtz instability acts 
to shed its outer later. However, at 0.75 Myr only 12\% of the mass of material 
that remains on the computational grid is found mixed into the hot wind. The bulk 
of the mass is still found in the cloud's elongated core and tail, and despite the 
driving of radiative shocks into the cloud(s) there is little increase in the 
temperature of the cloud material as it radiates heat, remaining cool and cohesive 
as it leaves the computational grid. 

\subsubsection{Fractal Cloud}\label{frac_cloud}

\begin{figure}[p]
\epsscale{0.8}
\plotone{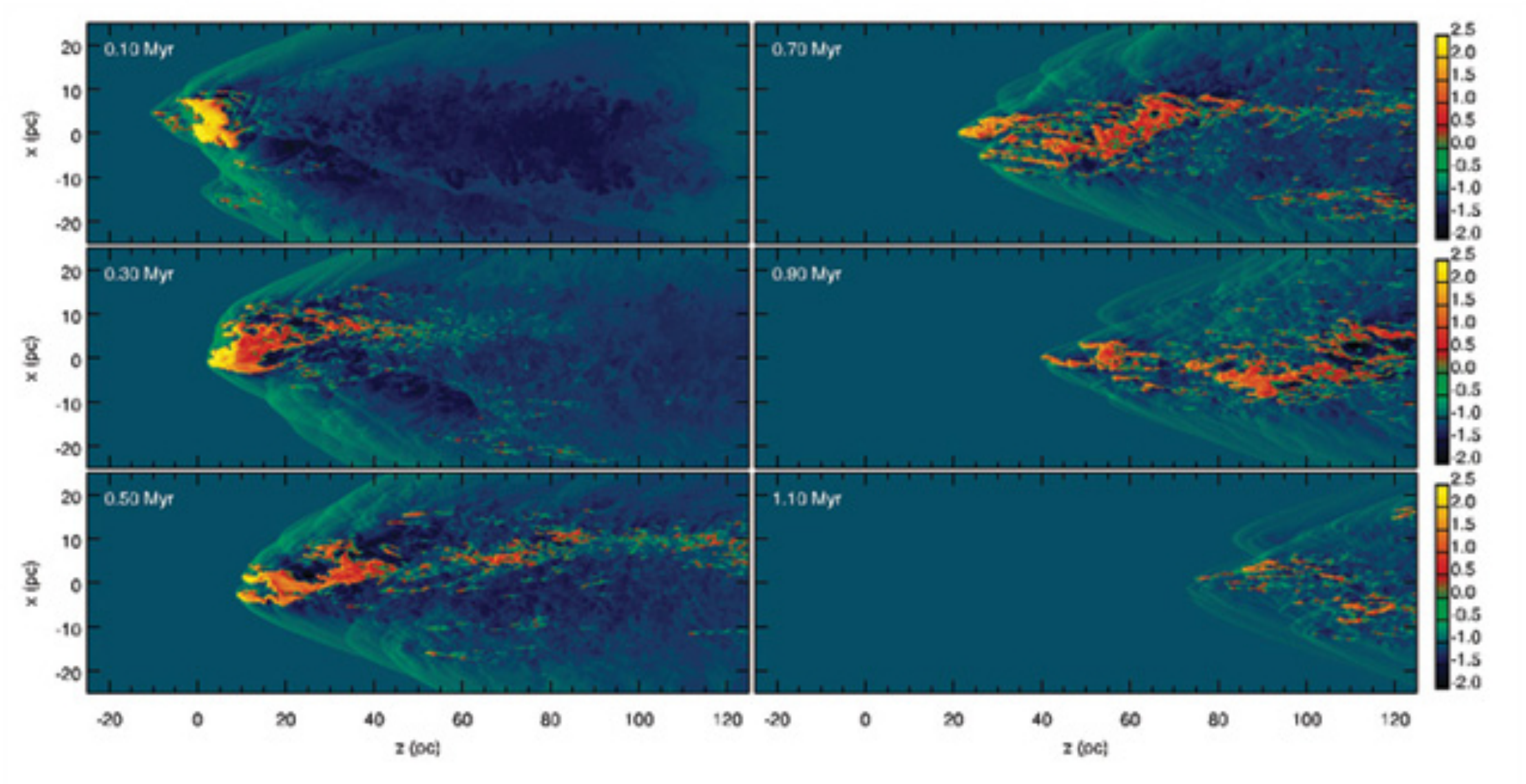}
\caption{Logarithm of the density through the y=0 plane in model rf384 showing
  evolution of a radiative fractal cloud.}\label{fig:dens_cld}
\end{figure}

\begin{figure}[p]
\epsscale{0.8}
\plotone{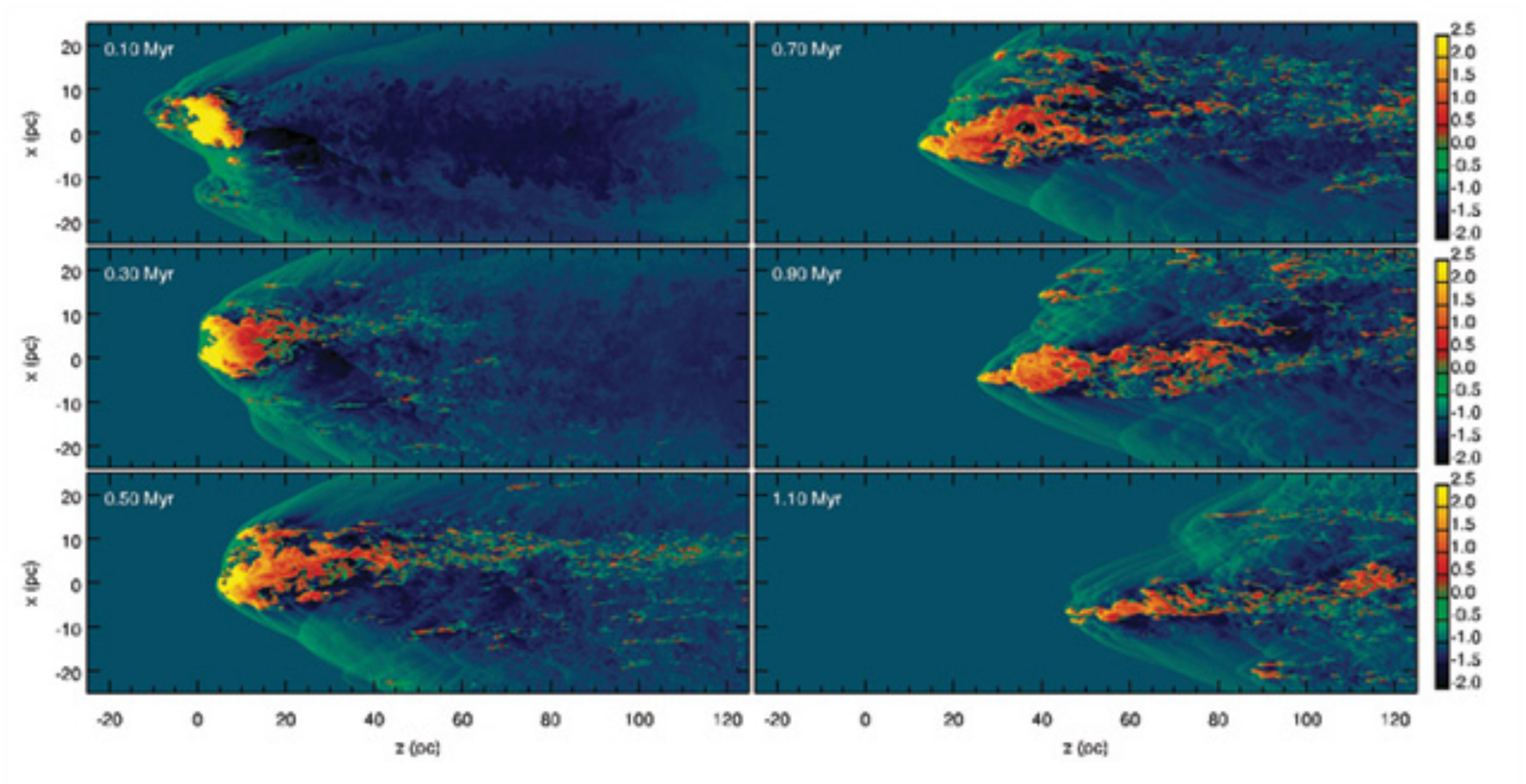}
\caption{Logarithm of the density through the y=0 plane in model rfd384 showing
  the effect of the clouds initial density on the
  evolution of a radiative fractal cloud.}\label{fig:dens_hd}
\end{figure}

Figure \ref{fig:dens_cld}  shows the evolution of the radiative fractal cloud in 
model rf384 from 0.1 to 1.1 Myr through density slices at intervals of 0.2
Myr. Figure \ref{fig:dens_hd} is identical to Figure \ref{fig:dens_cld}, but
shows the evolution of a cloud (rfd384) with twice the density and mass than 
rf384. As with the radiative spherical cloud, a bow shock immediately forms 
upstream of the cloud. However, a significant effect of the inhomogeneous 
structure of the cloud is the formation of a shock off each ridge on the cloud's 
surface that is exposed to the wind. This has the effect of creating a ``web'' of 
interacting shocks. 

\begin{figure}[tbp]
\epsscale{0.8}
\plotone{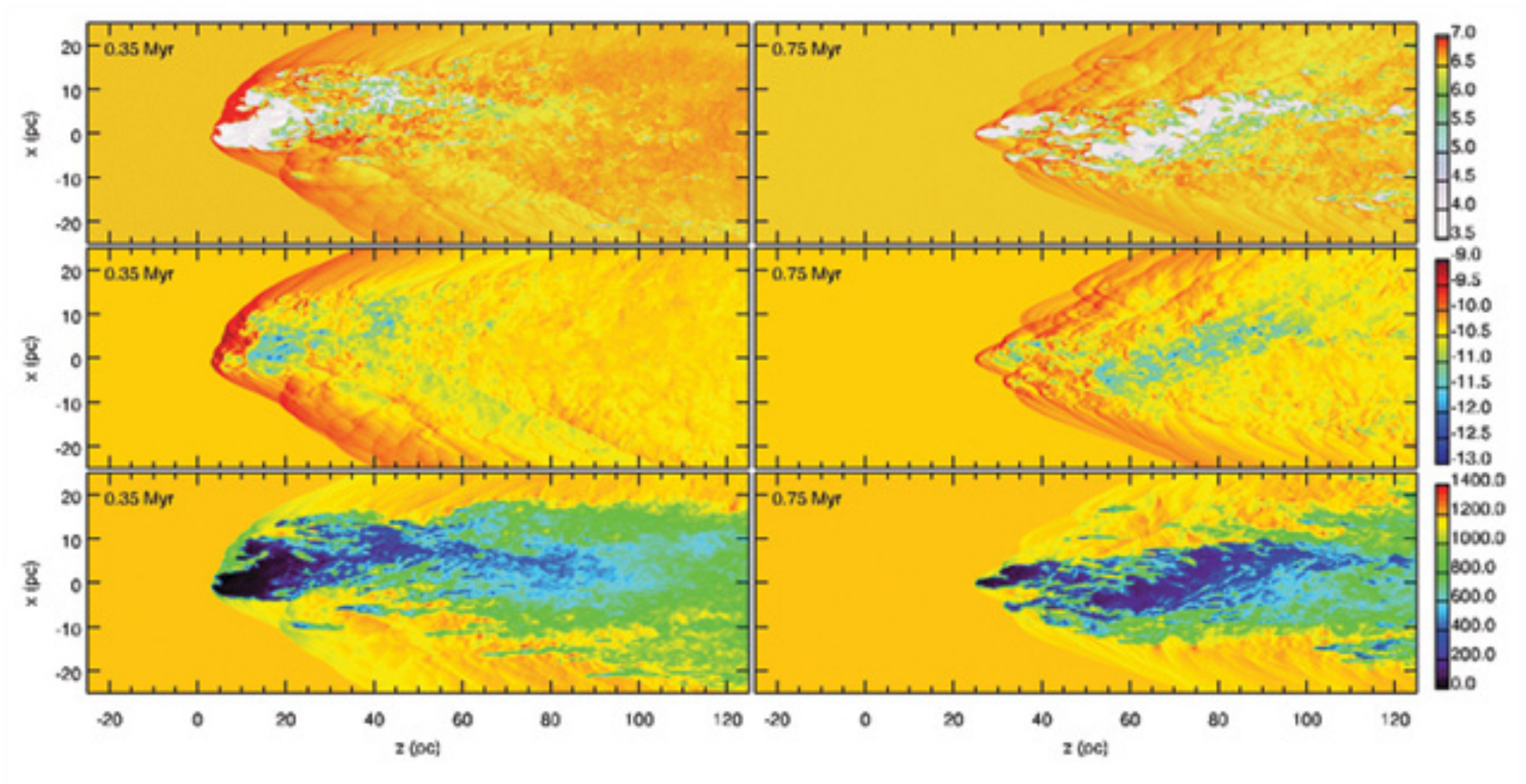}
\caption{Log-temperature (top), log-pressure (middle), and velocity (bottom)
  at 0.35 (right) and 0.75 (left) epochs through the y=0 plane in model
  rf384.}\label{fig:var_cld}
\end{figure}

By 0.3 Myr (Fig. \ref{fig:dens_cld}; middle left panel), the density at the
front of the cloud has jumped to $n \sim 1000 ~\rm cm^{-3}$, but fallen to
approximately $n = 10 ~\rm cm^{-3}$ at the rear of the cloud. Despite the high 
pressure exerted on the cloud, very little heating occurs, with the cloud 
maintaining temperatures of $T \sim 10^3 - 10^4 ~\rm K$ (Fig. \ref{fig:var_cld}; 
left upper two panels). As the cloud evolves (Fig. \ref{fig:dens_cld}; right
panels), small cloudlets are broken off the main cloud by the Kelvin-Helmholtz
instability. These cloudlets form a filamentary structure downstream of the
main cloud and have velocities in the range of $0 - 400 ~\rm km~s^{-1}$,
somewhat slower than the velocity of the surrounding stream ($v > 800 ~\rm
km~s^{-1}$) (Fig. \ref{fig:var_cld}; bottom panels). 

The cloud is exposed to the high temperature and velocity of the wind at all 
points along the front surface of the cloud. The cloud breaks up fastest in 
regions where the density is lowest and the radius of the cloud is at its 
smallest. The cloud is continually eroded by the Kelvin-Helmholtz instability, the 
fragments of which are immersed in a low pressure, turbulent gas. At 0.75 Myr, the 
percentage of the mass of cloud material remaining on the computational 
grid that has mixed into the hot wind is $\sim$ 25\%. The bulk of the original
cloud mass is found in the remaining core remnant and the stream of small,
dense $\sim$ 1 pc cloudlets. If these cloudlets are exposed to the wind they
produce their own high pressure bow shock upstream of their position. However,
the majority of the cloudlets are sheltered from the wind by other fragments
broken off from the main cloud. These cloudlets survive to leave the computational 
grid.

The higher density of the cloud in model rfd384 (Fig.  \ref{fig:dens_hd}) results in the cloud 
retaining is structural integrity far  longer than the lower density cloud in rf384. While the
evolution  of the higher density cloud is overall very similar to that in rf384, the break-up 
of the cloud via the Kelvin-Helmholtz instability  is slower: At 1.1 Myr (bottom right 
panel), a significant bulk of the cloud material  still remains on the computational grid and 
displays a similar morphology to  the cloud in rf384 at 0.7 Myr (Fig. \ref{fig:dens_cld}; top
right panel). In  contrast to rf384, at 0.75 Myr only 11\% of the clouds
mass remaining on  the computational grid is found mixed into the hot
wind. However, it is  likely this amount will increase as the cloud is further
eroded by the  Kelvin-Helmholtz instability.

\begin{figure}[tbp]
\epsscale{0.8}
\plotone{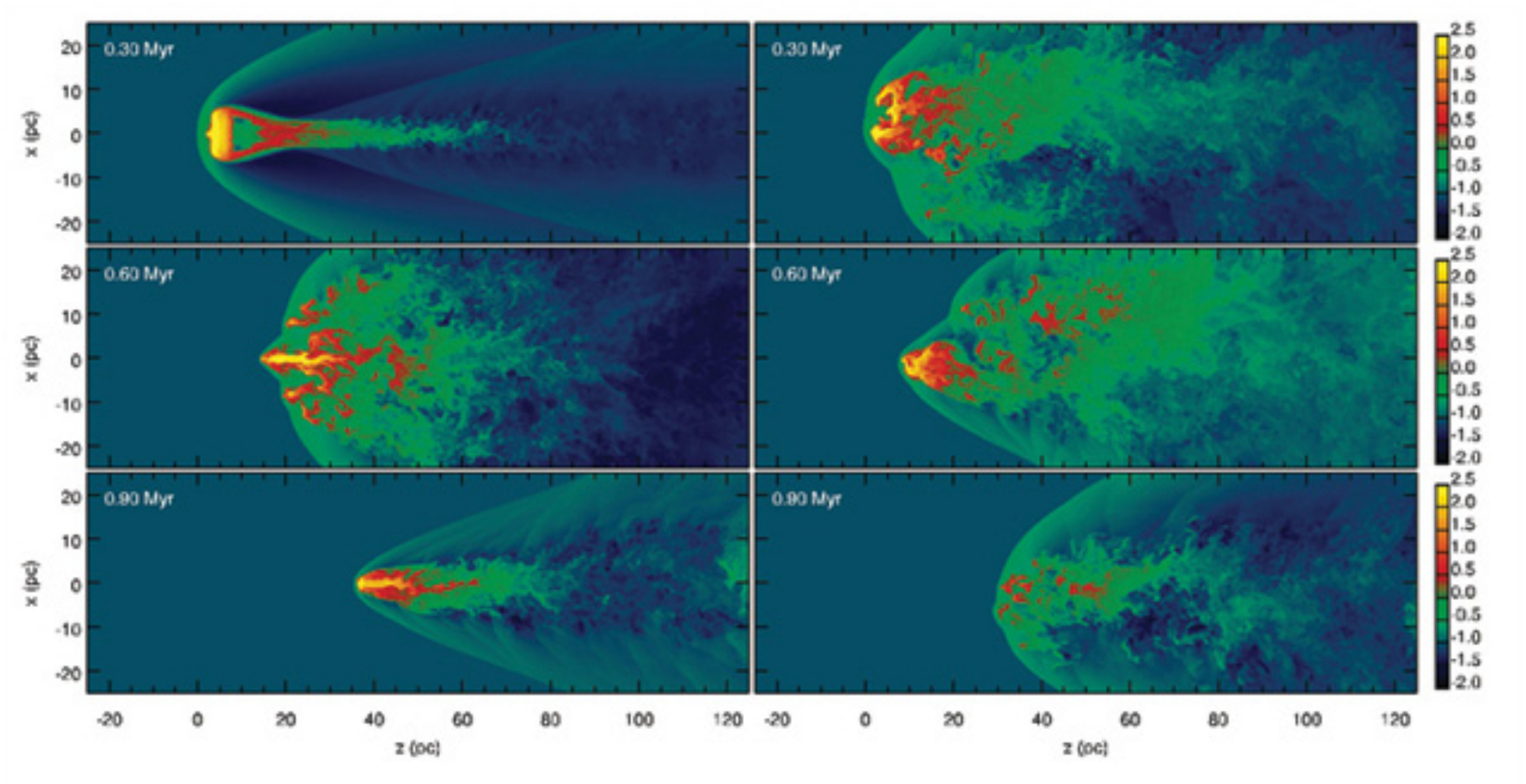}
\caption{Logarithm of the density through the y=0 plane in models as384 (left)
  and af384 (right) showing the evolution of an adiabatic spherical and
  fractal cloud respectively.}\label{fig:dens_adi}
\end{figure}

As observed by \citet{Xu1995}, the initial shape of the cloud has an effect on
its subsequent evolution, with our fractal cloud fragmenting faster than the
spherical cloud. While this is due in part to the lower average density of the
fractal cloud,  even the high density fractal cloud, whose average density
and total mass is greater than that of the spherical cloud (see Table
\ref{sim_param}), has a greater degree of fragmentation and is less cohesive
when it leaves the computational grid. This is a result of the inhomogeneous
nature of the fractal cloud's initial density distribution and the larger
cross-section is presents to the incident wind (Fig. \ref{fig:dens_profile};
left panel). The cloud first fragments along regions where the wind finds paths of
least resistance, i.e. regions of low density. As a result the fractal cloud
breaks-up into multiple core fragments. As the wind finds no regions of least
resistance in the spherical cloud, it is able to retain a single cohesive 
structure for a longer period of time. Thus, not only is the initial geometry of 
the cloud important in determining its evolution, the distribution of the cloud's 
density determines how quickly the cloud begins to fragment. More homogeneous 
density distributions would result in less initial fragmentation.   

\begin{figure}[tbp]
\epsscale{0.8}
\plotone{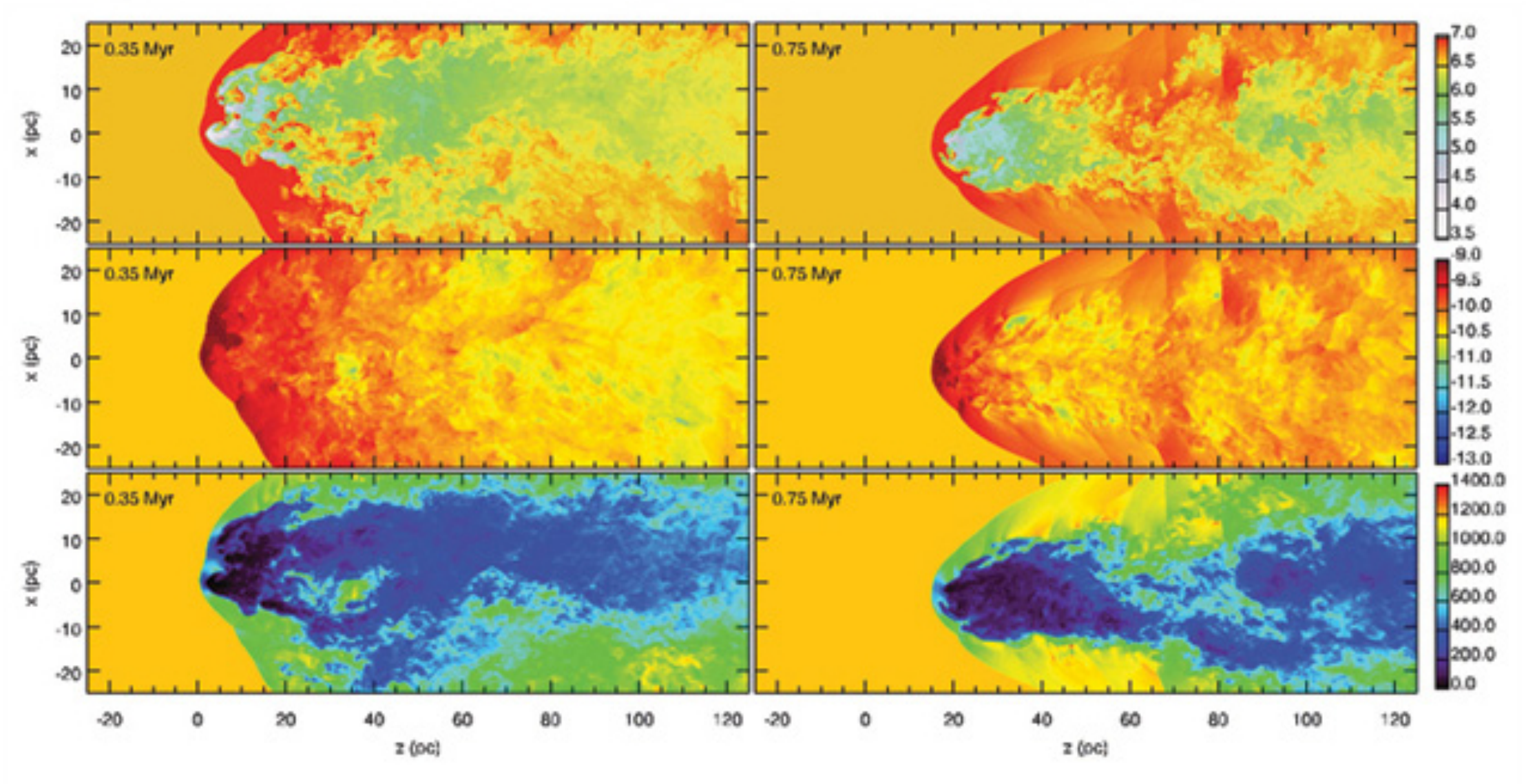}
\caption{Log-temperature (top), log-pressure (middle), and velocity (bottom)
  at 0.35 (right) and 0.75 (left) epochs through the y=0 plane in model
  af384.}\label{fig:var_af}
\end{figure}

\subsection{Effect of Radiative Cooling}

\subsubsection{Adiabatic Case}

In order to understand the degree as to which the inclusion of radiative
cooling affects the survival of a cloud, we performed 2
simulations in which cooling was neglected: as384 and af384 for the spherical
and fractal case respectively. Figure \ref{fig:dens_adi} shows the density of
the adiabatic clouds in as384 (left) and af384 (right) at 0.3, 0.6, 0.9 Myr
epochs. It can be immediately seen that in the absence of radiative cooling
there is a far greater degree of mixing of cloud material with the surrounding
stream, with 40\% and 59\% of the fractal and spherical clouds' masses 
respectively on the computational grid found mixed into the hot wind at 0.75 Myr. 

The destruction of the adiabatic fractal cloud occurs faster and is more
complete than the adiabatic spherical cloud. At 0.9 Myr the cloud has been almost 
completely destroyed (Fig. \ref{fig:dens_adi}; bottom right panel), with the 
cloud material mixed into the hot gas. The initial interaction with the wind is 
the same as in the radiative case, with a bow shock forming upstream of the cloud. 
However, the cloud gas quickly begins to heat to temperatures of the order of $T 
\sim 10^6 ~\rm K$ (Fig. \ref{fig:var_af}; upper left panel) and the cloud
expands. Again the cloud breaks up first in regions where the density is low
and the cloud radius is at its smallest. The high pressure exerted on the
cloud (Fig. \ref{fig:var_af}; middle left panel) and the Kelvin-Helmholtz
instability act to strip material from the cloud. While this material is
able to survive in the radiative model, here the cloudlets broken
off the main cloud are quickly heated and destroyed. The bulk velocity of the
gas downstream of the main cloud is lower than that found in the radiative case
(Fig. \ref{fig:var_af}; lower panels). 

\begin{figure}[tbp]
\epsscale{0.8}
\plotone{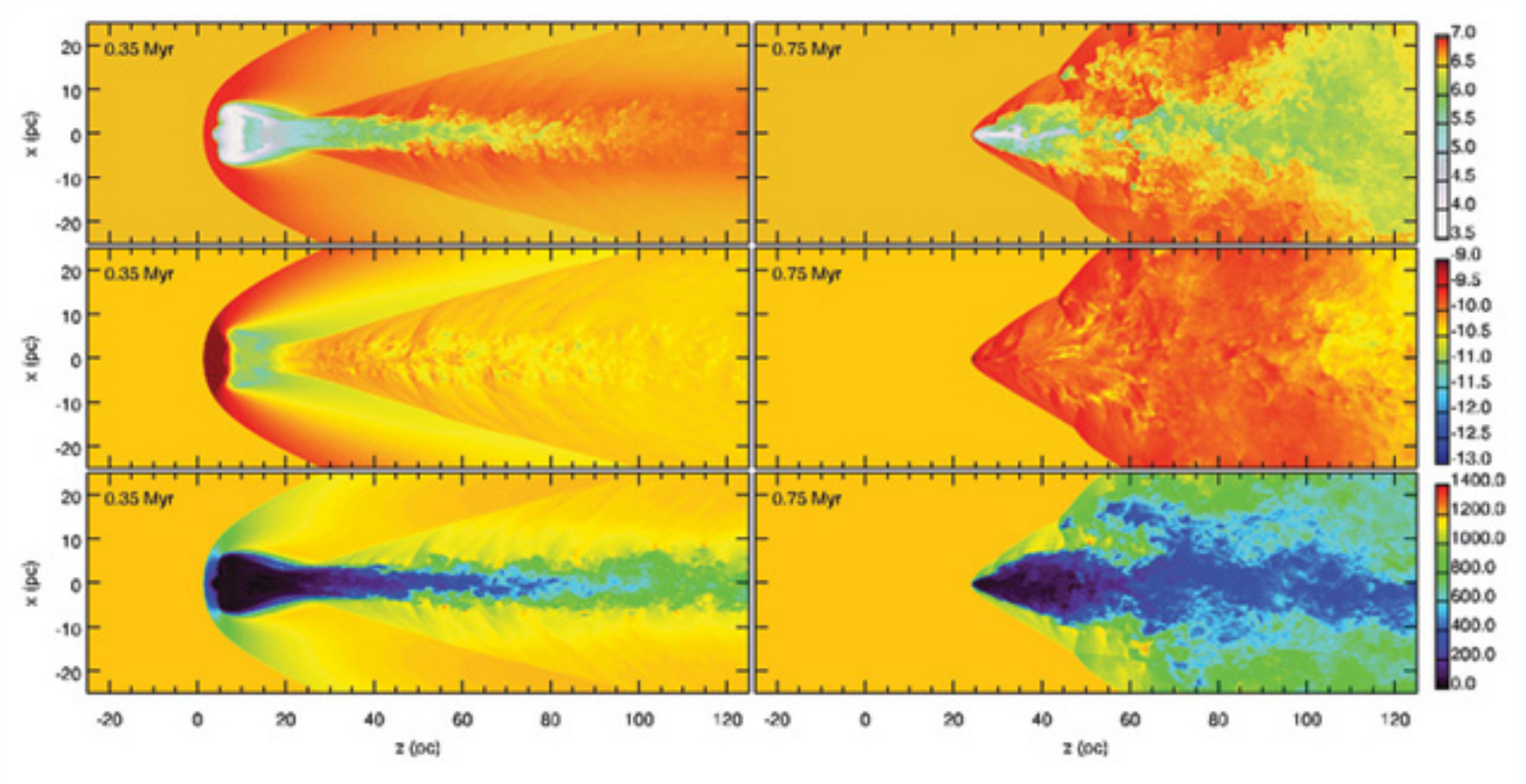}
\caption{Log-temperature (top), log-pressure (middle), and velocity (bottom)
  at 0.35 (right) and 0.75 (left) epochs through the y=0 plane in model
  as384.}\label{fig:var_as}
\end{figure}

In the case of the adiabatic spherical cloud, the shedding of the cloud's outer 
layer by the Kelvin-Helmholtz instability, also seen in the radiative model, is 
greatly enhanced. By 0.9 Myr (Fig. \ref{fig:dens_adi}; bottom left panel), only 
the cloud core remains and is subsequently destroyed by the Kelvin-Helmholtz
instability as the simulation progresses. As with the fractal cloud, the bulk 
velocity of the gas downstream of the main cloud is lower than that of the
radiative spherical cloud at the same time (Fig. \ref{fig:var_as}; lower
right panel). The evolution of a spherical adiabatic cloud is discussed in 
more detail below.  

\subsubsection{Cloud Survival}\label{survival}

\begin{figure}[tbp]
\epsscale{0.8}
\plotone{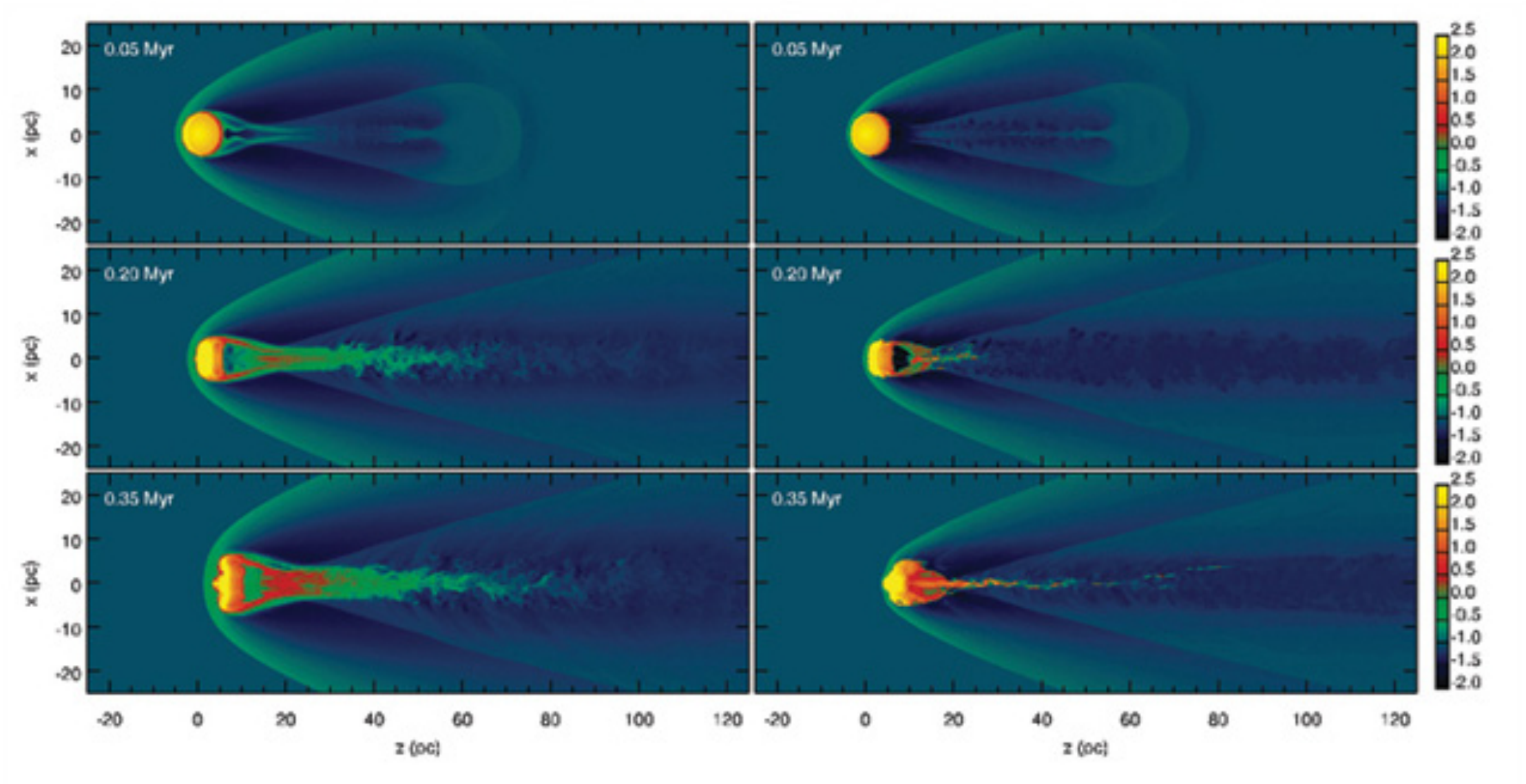}
\caption{Logarithm of the density through the y=0 plane in models as384 (left)
  and rs384 (right) showing the initial evolution of a spherical cloud over
  the first 0.35 million years.}\label{fig:dens_initial}
\end{figure}

One of the significant effects of the inclusion of radiative cooling is the
longer life time of the impacted cloud. While this effect has been observed in
the past by other authors \citep[e.g.][]{Mellema2002,Melioli2005}, our study
is of higher resolution and does not assume any symmetry, making it is a
useful exercise to directly compare the simple case of the evolution of the 
spherical cloud in both our adiabatic (as384) and radiative (rs384) models in 
order to determine the mechanism behind the radiative cloud's survival. The initial
interaction of the wind and cloud is shown via density slices in Figure 
\ref{fig:dens_initial} at 0.05, 0.20 and 0.35 Myr epochs in both adiabatic (left) 
and radiative (right) models. It can be clearly seen that while the evolution 
begins almost identically with a bow shock forming upstream of the cloud, in the 
adiabatic case cloud material immediately starts being ablated from the back of 
the cloud.

As in the radiative model, a shock propagates through the adiabatic cloud. 
However, the initial density increase observed in the cloud is not as extreme 
(e.g. $n < 1000 ~\rm cm^{-3}$). The cloud material is heated to temperatures of 
the order of $T > 10^5 ~\rm K$ and the cloud expands. The shock travels though the 
cloud, reflecting off the back surface at approximately 0.27 Myr, somewhat faster 
than in the radiative cloud. The adiabatic cloud expands transversely as it is 
accelerated downstream. However, this transverse expansion is suppressed in the 
radiative cloud as a result of the lower degree of heating of the cloud gas (Fig.
\ref{fig:var_sph}; upper panels). In both cases, the Kelvin-Helmholtz instability 
acts to strip material from the edges of the cloud forming a tail of material
downstream of the cloud position (Fig. \ref{fig:dens_initial}; middle and lower 
panels). In the adiabatic model, this tail is geometrically thick with density and 
temperature $n \sim 1 ~\rm cm^{-3}$ and $T \sim 10^6 ~\rm K$ respectively. In 
contrast, the tail formed in the radiative model is geometrically thin with 
density $n \sim 10 ~\rm cm^{-3}$ and temperature $T \sim 10^4 ~\rm K$.

In the adiabatic model, the internal cloud shock reflects again off the front the
cloud at approximately 0.35 Myr. At this time, the transverse expansion of the
cloud persists and the Kelvin-Helmholtz instability continues to strip
material from the clouds exterior into the downstream tail of gas. (Fig.
\ref{fig:dens_initial}; lower left panel). The transverse expansion (Fig.
\ref{fig:var_as}; middle panels) results in a higher rate of acceleration in
the adiabatic model. As a consequence, the cloud has a lower relative Mach
number relative to the stream (Fig. \ref{fig:var_as}; lower left panel) than
the radiative cloud (Fig. \ref{fig:var_sph}; lower left panel). The growth
rate of the Kelvin-Helmholtz instability is lower for higher Mach numbers and
its effect is strongly diminished for the radiative cloud. The adiabatic cloud
is more easily disrupted and destroyed. This can be dramatically seen in the
middle left panel of Figure \ref{fig:dens_adi} where the Kelvin-Helmholtz
instability has stripped the entire outer layer of the adiabatic spherical
cloud \citep[see also Fig. 3 of][]{Orlando2005}.

Since a radiative cloud is broken-up via the Kelvin-Helmholtz instability into
a filamentary structure of small $\sim$ 1 pc sized clouds, the survival of these 
small clouds is of interest. We now compare the cloud crushing time ($t_{\rm 
crush}$) and the Kelvin-Helmholtz timescale ($t_{\rm KH}$) to the cooling time 
($t_{\rm cool}$) of a cloud with radius $R_{\rm c} = 1 ~\rm pc$, density 
$\rho_{\rm c} = 10 ~\rm cm^{-3}$, temperature $T_{\rm c} = 10^4 ~\rm K$, and 
velocity $v_{\rm c} = 200 ~\rm km~s^{-1}$. The cloud crushing time of such a cloud 
is $t_{\rm crush} \approx R_{\rm c}/v_{\rm sh} \approx (\rho_{\rm c}/\rho_{\rm w}) 
R_{\rm c}/v_{\rm w} = 3 \times 10^{12} ~\rm s$, where the density and velocity of 
the wind is  $\rho_{\rm w} = 0.1 ~\rm cm^{-3}$ and $v_{\rm w} = 1000 ~\rm 
km~s^{-1}$ respectively. The Kelvin-Helmholtz timescale is $t_{\rm KH} = R_{\rm 
c}(\rho_{\rm c} + \rho_{\rm w})/(v_{\rm c}-v_{\rm w})(\rho_{\rm c} \rho_{\rm w}) = 
3 \times 10^{11} ~\rm s$. The cooling time for a 1 pc cloud in our simulations is 
of the order of $10^{10}$ seconds, somewhat shorter than the cloud crushing time 
and the Kelvin-Helmholtz timescale, suggesting that the cloudlets may remain
sufficiently stable to ablation and survive to later times. In addition, we
note that the self-gravity of the clouds may cause them to collapse, becoming
more difficult to disrupt \citep{Mellema2002}.

\begin{figure}[tbp]
\epsscale{1.0}
\plottwo{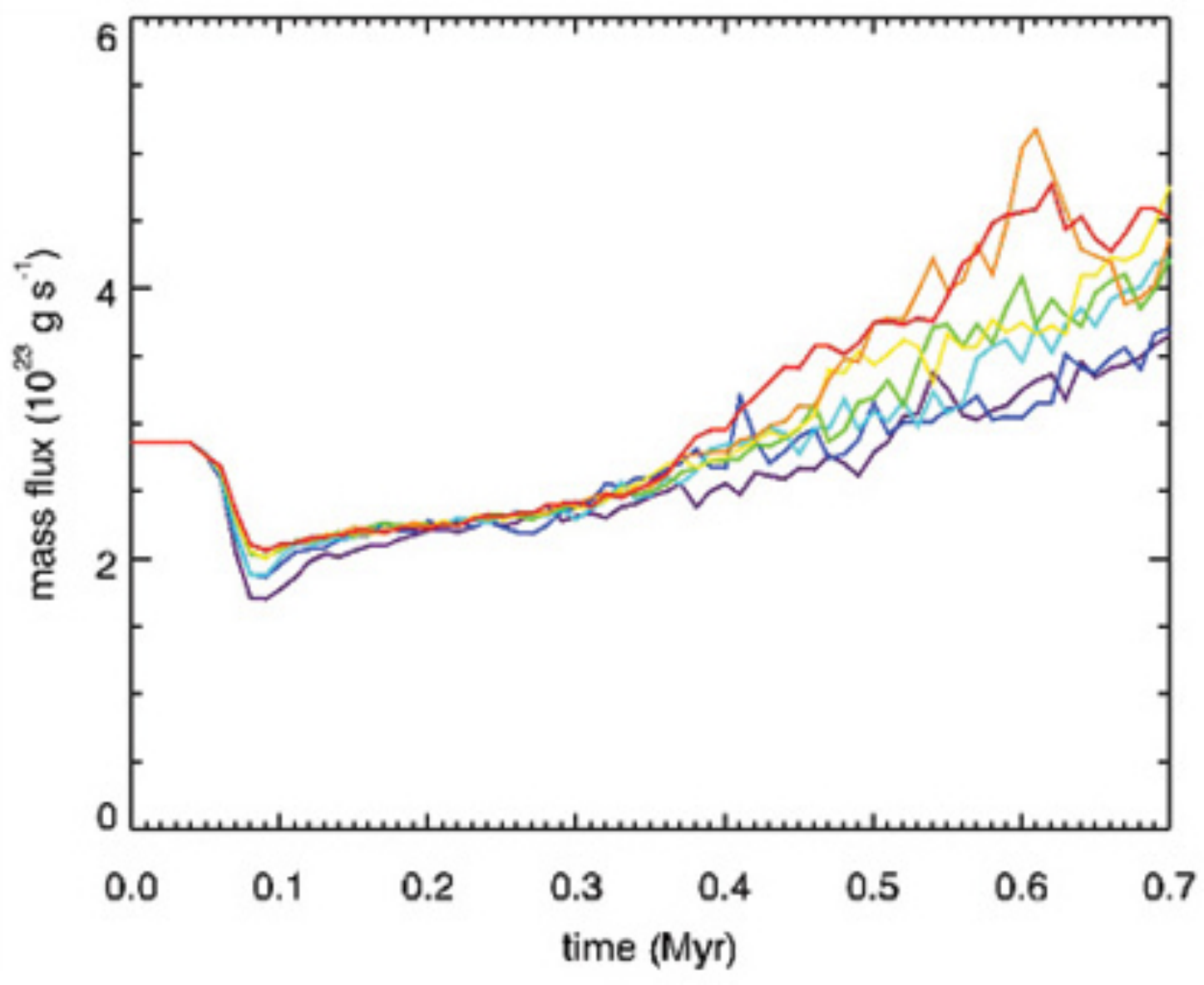}{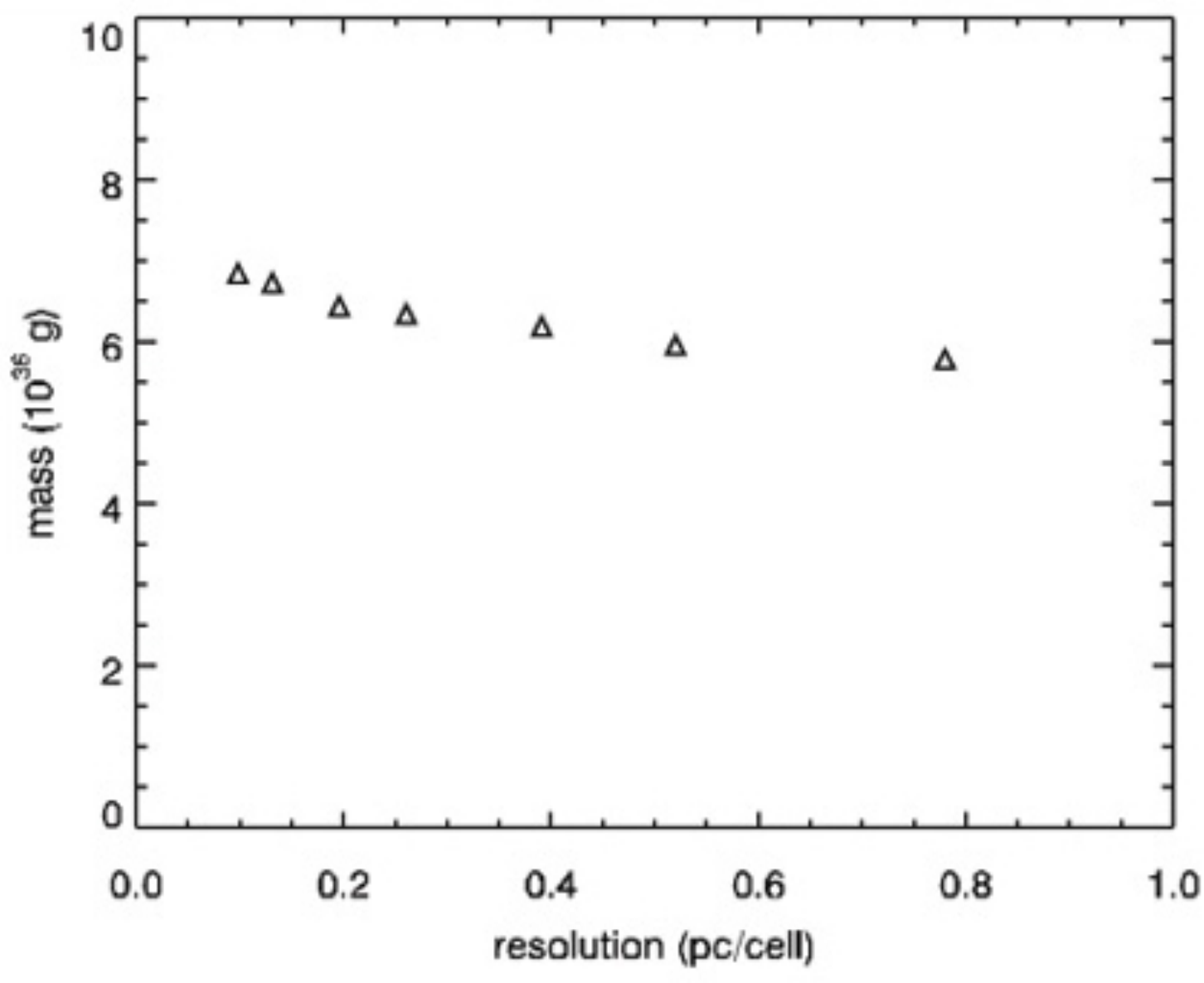}
\caption{Left: mass flux through a surface at z=0.75pc for each resolution (online:
  0.78 pc/cell [navy], 0.52 pc/cell [blue], 0.39 pc/cell [cyan], 0.26 pc/cell
  [green], 0.20 pc/cell [gold], 0.13 pc/cell [orange], and 0.10 pc/cell
  [red]). Right: Total mass flux integrated over the first 0.7 Myr as a
  function of the numerical resolution.}\label{fig:mass_flux}
\end{figure}

\section{RESOLUTION STUDY}\label{resolution}

\subsection{Mass Flux}

In order to test the dependence of our results on the numerical resolution of the 
code, we have performed seven simulations of the radiative fractal cloud interacting 
with a supersonic wind at increasing resolutions from 0.78 - 0.10 pc per cell 
width (see Table \ref{sim_param}). We calculated the flux of mass through a 
surface at $z = 75 ~\rm pc$ over the first 0.7 million years of each simulation in 
our resolution study. Figure \ref{fig:mass_flux} shows this mass flux for each 
simulation as a function of time (left panel), as well as the total mass flux 
integrated over the first 0.7 Myr of the simulation as a function of resolution 
(right panel). The mass flux over the first 0.4 Myr of the evolution is similar at 
all resolutions. This is a result of the well resolved hot stream of gas passing 
through the flux surface. After this point, the cloud material begins to pass 
though the surface and the mass flux starts to vary with resolution. 

\begin{figure}[tp]
\epsscale{0.5}
\plotone{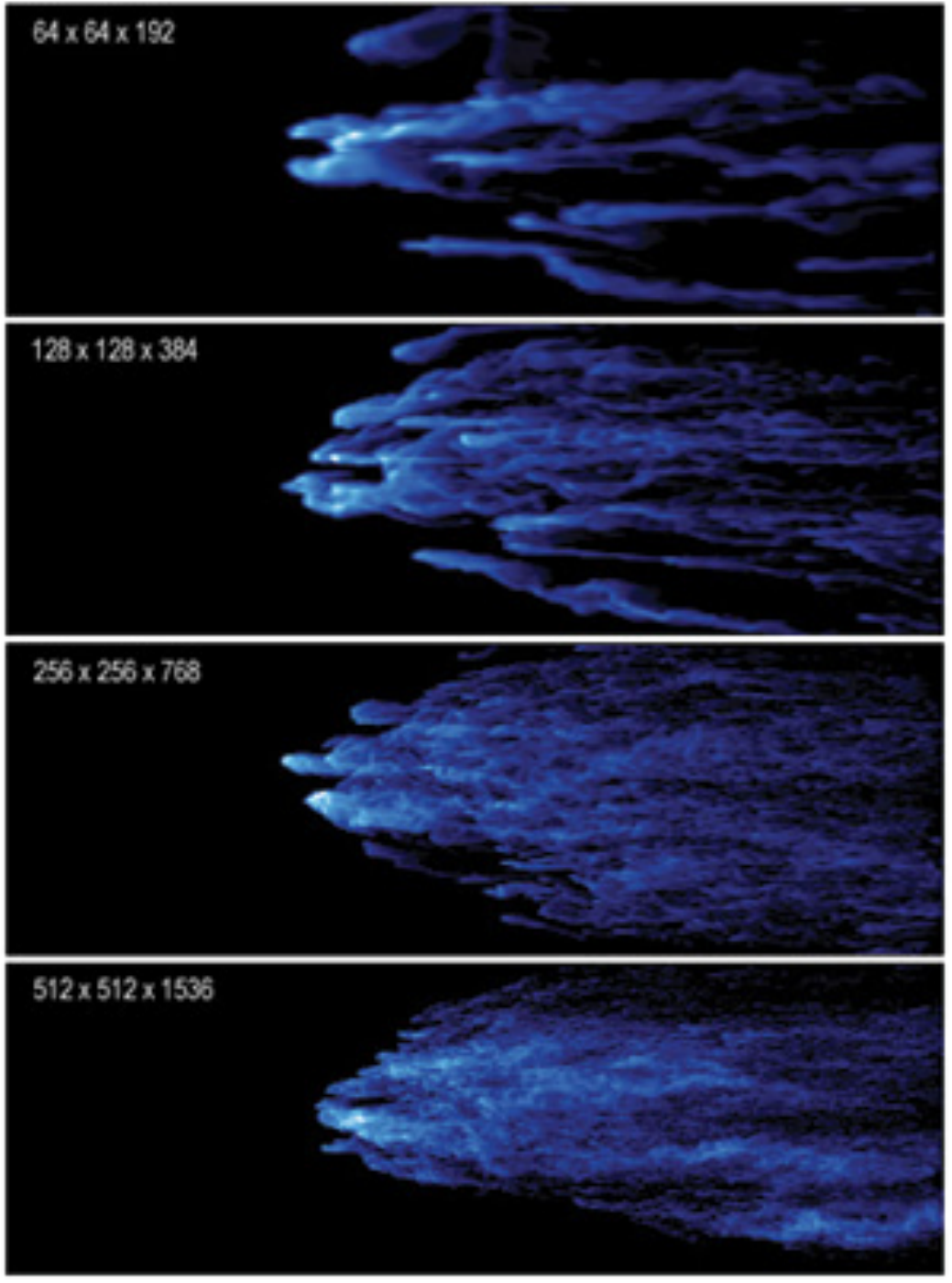}
\caption{Volume renderings of the projected density at increasing resolution (top 
to bottom: 0.78 pc/cell, 0.39 pc/cell, 0.2 pc/cell, and 0.1 
pc/cell).}\label{fig:res_dens}
\end{figure}

The initial drop in the mass flux seen in the left hand panel of Figure 
\ref{fig:mass_flux} is due to the rarefaction that passes though the flux surface 
(see Fig. \ref{fig:dens_sph}; top left panel). From approximately 0.2 to 0.4 
Myr, the density of the stream increases as mass is ablated from the rear of the 
cloud resulting in a similar flux at all resolutions. During this time, the mass 
flux gradually increases as the $n = 0.1 ~\rm cm^{-3}$ mass loaded stream of gas 
passes through the flux surface. At approximately 0.4 Myr, the mass flux begins to 
vary rapidly, dramatically increasing in each simulation. The large variation in 
the mass flux is due to the turbulent nature of the gas passing though the flux 
surface and the dense cloudlets immersed within this gas. The cloudlets pass 
through the surface at different times resulting in fluctuations in the mass flux. 

\begin{figure}[tbp]
\epsscale{0.7}
\plotone{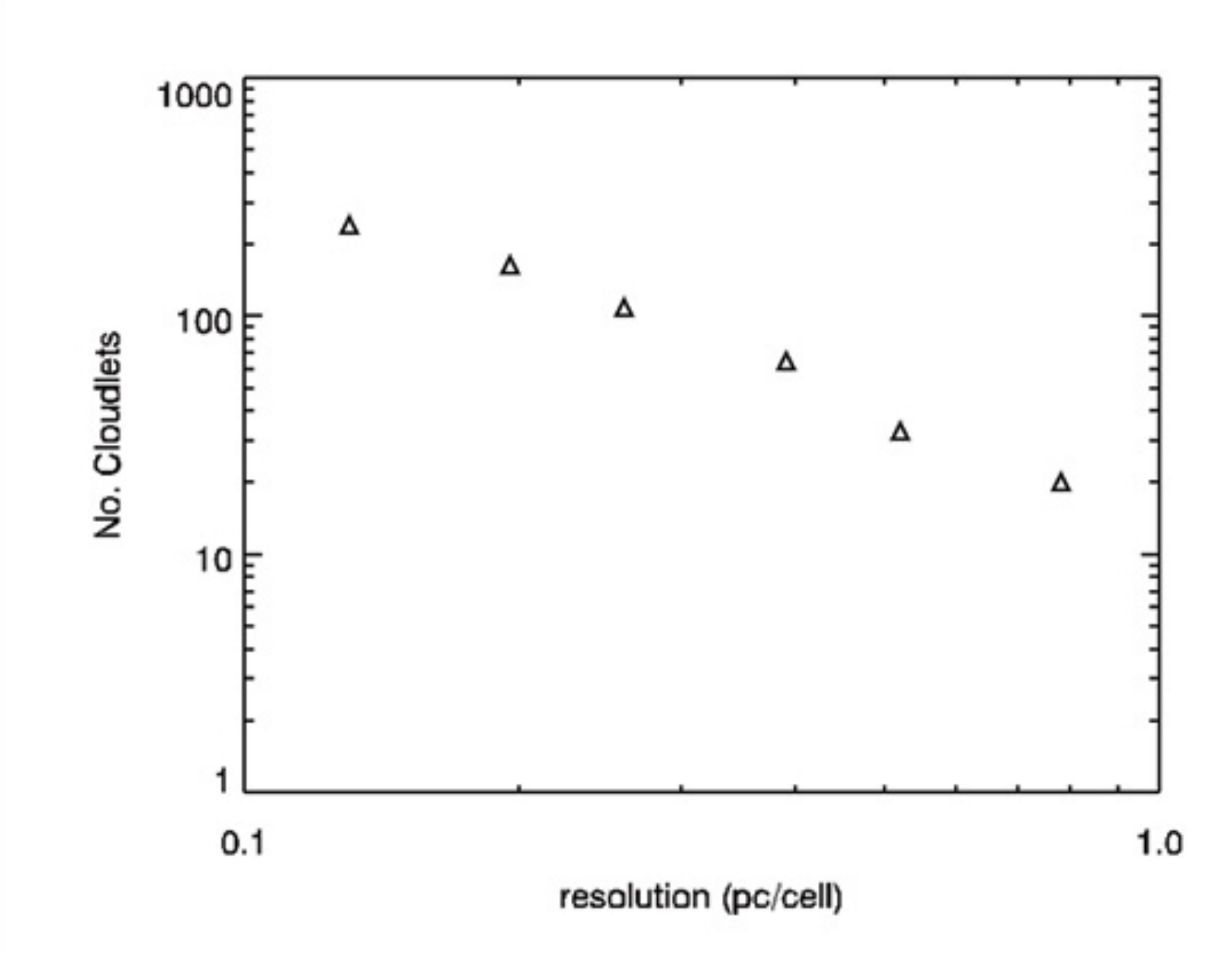}
\caption{Number of cloudlets formed via the break-up of the radiative fractal
  cloud at 0.7 Myr as a function of numerical resolution.}\label{fig:number_clouds}
\end{figure}

There is a general trend of increasing mass flux with the resolution of the 
simulation. This trend can be explained by the increase in fragmentation of the 
cloud with increasing numerical resolution. Figure \ref{fig:res_dens} shows volume 
renderings  of the projected density at 0.7 Myr in models rf064, rf128, rf256 and 
rf512, which have resolutions of 0.78, 0.39, 0.20, and 0.10 pc per cell width 
respectively. The increase in the fragmentation of the cloud with resolution can 
clearly be seen, and will be discussed in more detail in \S~\ref{fragment}. At 
the highest resolution attempted in this study (0.1 pc), the filaments resemble a 
``foam'' of cloudlets, while at low resolution the cloud has been broken-up into 
only a few large fragments. As a consequence, the cross-section of dense material 
that passes through the flux surface at any given time after 0.4 Myr is larger in 
the higher resolution simulations.

The total integrated mass flux passing through the surface at $z = 0.75 ~\rm pc$ 
over the first 0.7 Myr also increases with numerical resolution (Fig. 
\ref{fig:mass_flux}; right panel). Again this is caused by the larger degree of 
fragmentation at high resolution, resulting in more ablation of cloud material 
from the back of the cloud. Between our highest and lowest resolution simulations 
the difference in mass flux is approximately 10\%. This discrepancy is likely to 
increase at higher resolutions, although it is possible that convergence may occur 
at extremely high resolution simulations ($> 0.1~\rm pc$) that utilize an adaptive 
mesh. For the 2 pc per cell width resolution of our global simulations in
paper {\sc i}, this error is increased to approximately 20\%. We will
discuss the impact of the numerical resolution on the results of paper
{\sc i} in \S~\ref{SB_winds}. 

\begin{figure}[tbp]
\epsscale{0.75}
\plotone{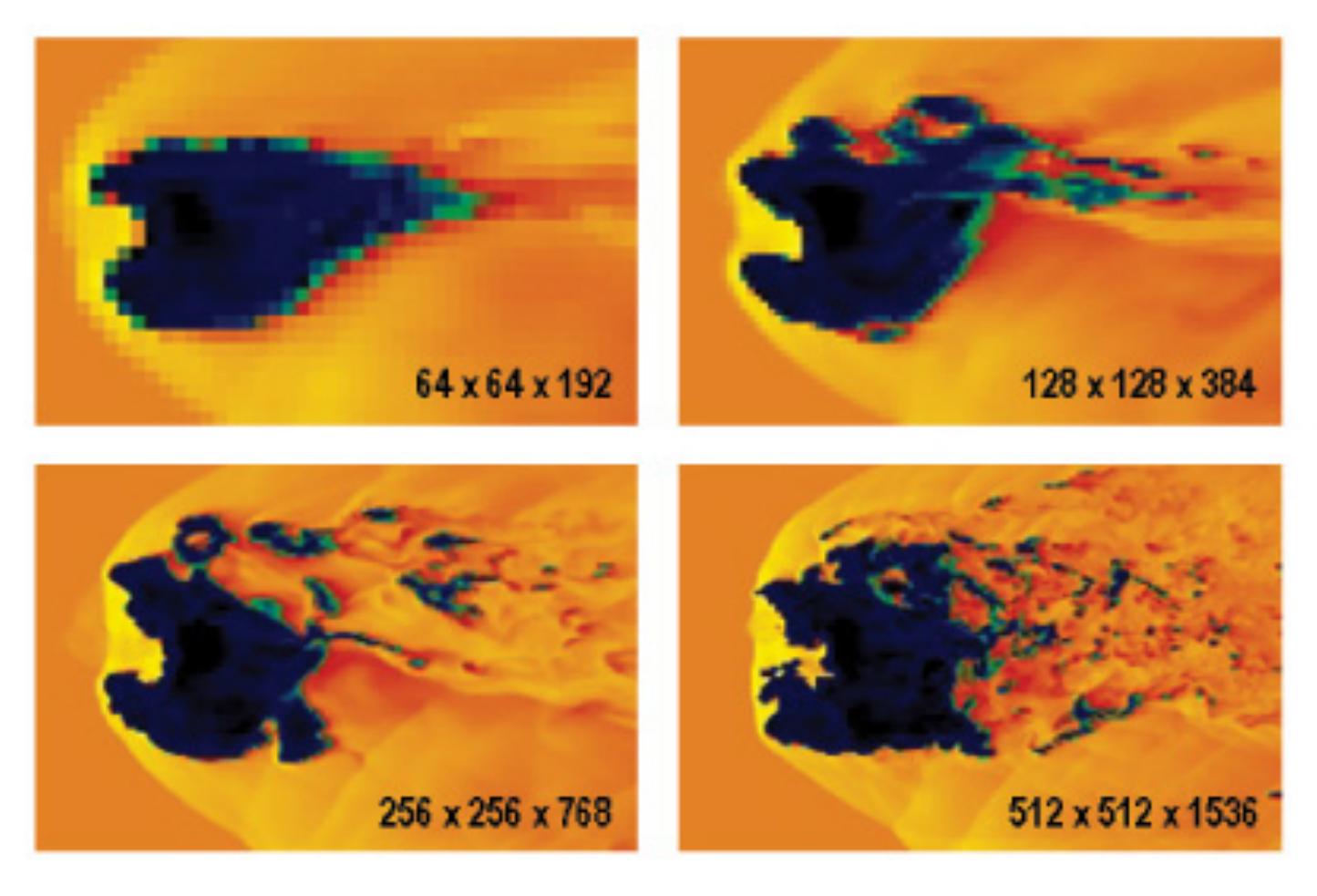}
\caption{Logarithm of the temperature (K) through the y=0 plane at 0.2 Myr of the 
radiative fractal cloud at 4 resolutions: 0.78 pc/cell (top left), 0.39 pc/cell 
(top right), 0.20 pc/cell (bottom left), and 0.10 pc/cell (bottom right). The 
cloud has been highlighted to show the effect of the Kelvin-Helmholtz 
instability.}\label{fig:Kelvin-helmholtz}
\end{figure}

\subsection{Cloud Fragmentation}\label{fragment}

The most significant effect of increasing the numerical resolution of the
simulation is the increase in fragmentation of the cloud. The increase in cloud
fragmentation can clearly be seen in Figure \ref{fig:res_dens}, with the the
cloud broken into only a few large fragments in the lowest resolution simulations, 
but 100's of fragments at higher resolution. Nevertheless, filamentary structure, 
where the concentration of cloudlets is higher, can still be made out at high 
resolution. These ``filaments'' are located in similar positions to the filaments 
in the low resolution simulations. 

We are able to calculate the properties of each cloudlet by using an algorithm 
which allows us to pick out and select fragments . Note that we impose a
minimum mass of $M_c = 10^{-3} ~\rm M_{\odot}$ for a fragment to be
selected. Figure \ref{fig:number_clouds} shows the number of cloudlets
produced as a function of numerical resolution. There is a general trend from
the number of cloudlets produced in the interaction to increase as a power law
with increasing resolution. Even with larger
computational resources and an adaptive mesh, this trend is likely to
continue ad infinitum. The dependence of the degree of fragmentation on the 
numerical resolution has been observed by other authors in both two-dimensional
\citep{Klein1994} and three-dimensional \citep{Stone1992} simulations of a
spherical cloud. This effect can be explained by the growth rate of
Kelvin-Helmholtz instability, which is faster at smaller wavelengths. As the
resolution is increased, this instability is increasingly resolved and more
fragmentation of the cloud is observed. This is illustrated in Figure
\ref{fig:Kelvin-helmholtz}, where the Kelvin-Helmholtz instability can clearly
be seen to increasingly fragment the cloud at higher numerical resolution. 

\begin{figure}[tbp]
\includegraphics[scale=0.3]{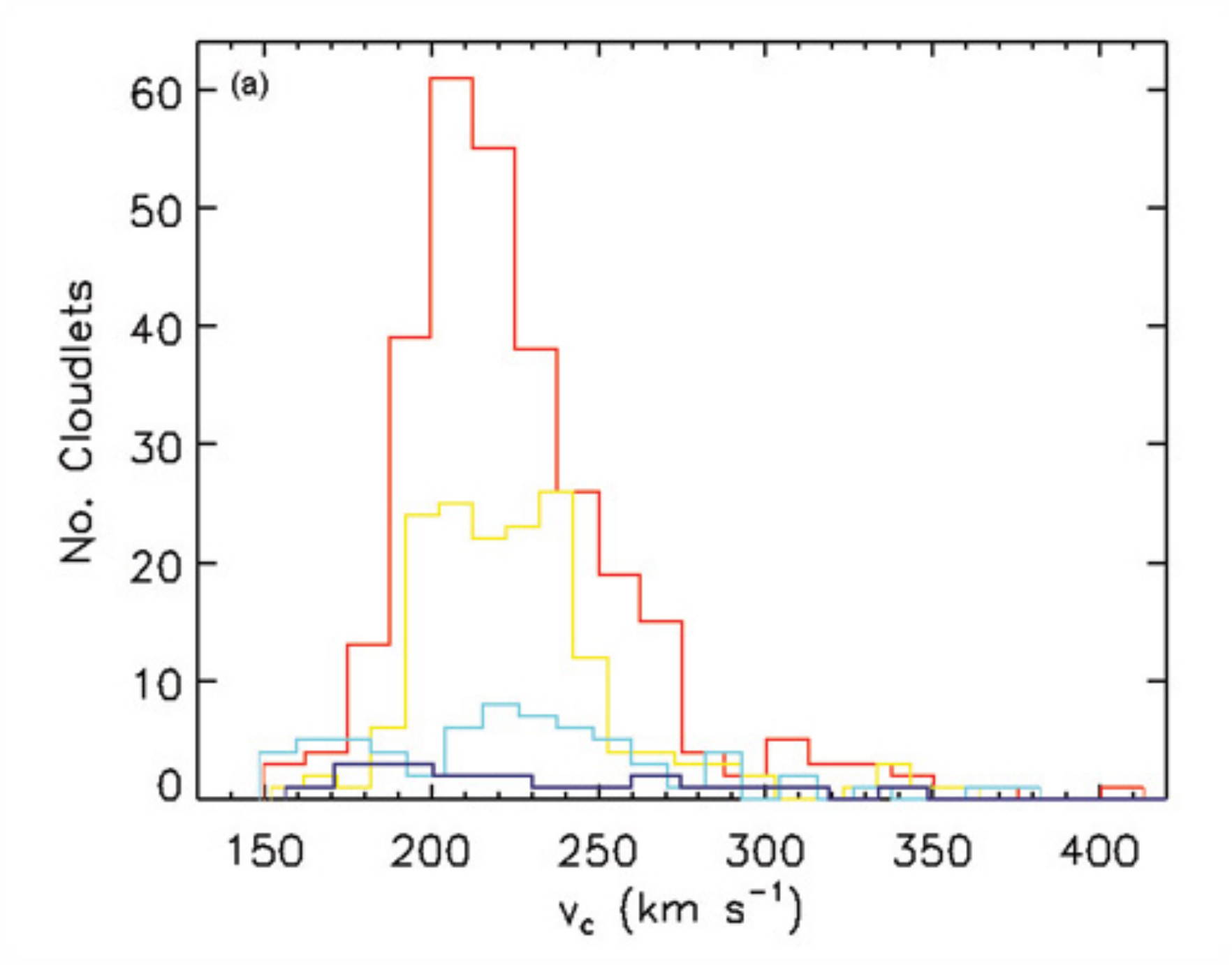}
\includegraphics[scale=0.3]{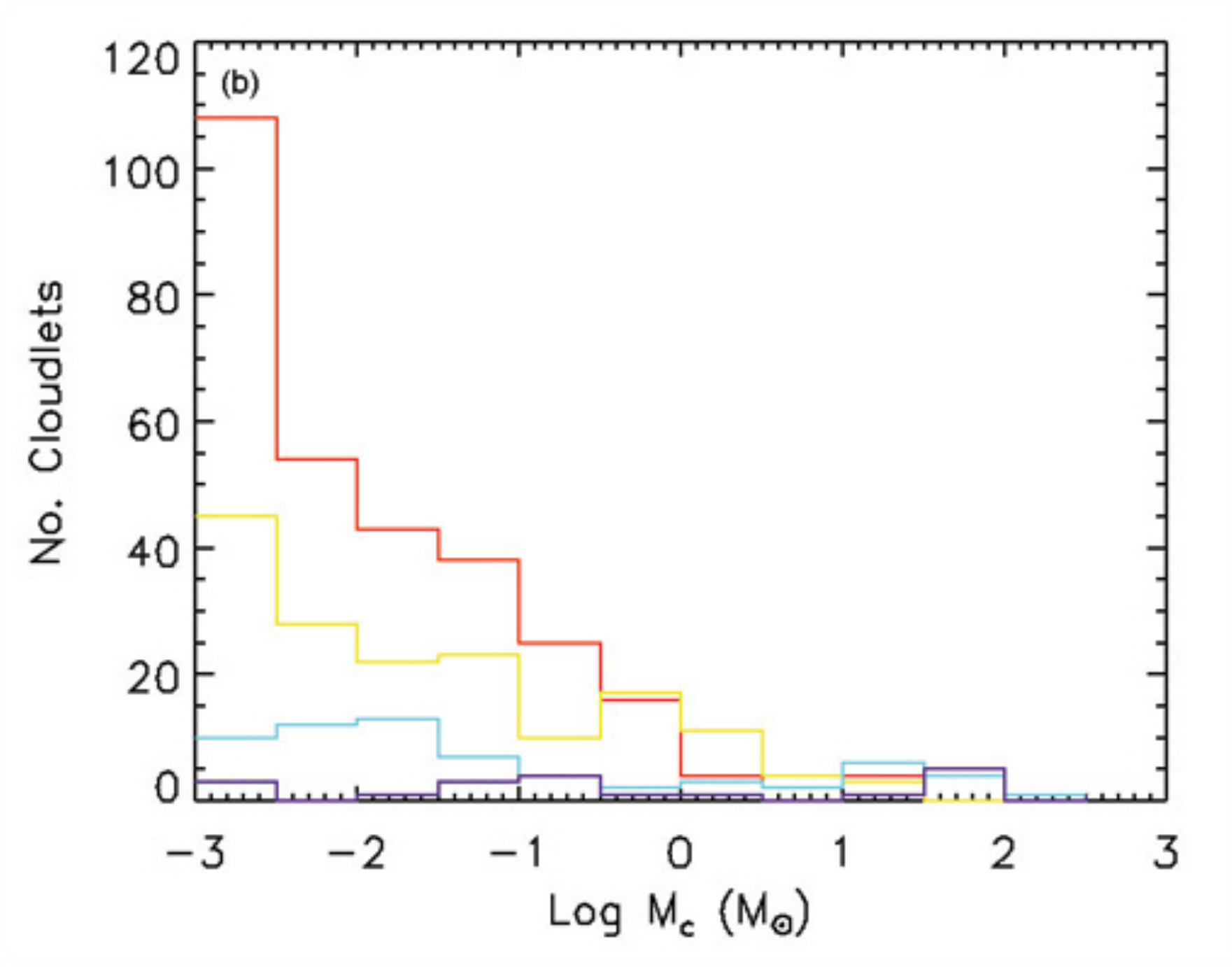}
\includegraphics[scale=0.3]{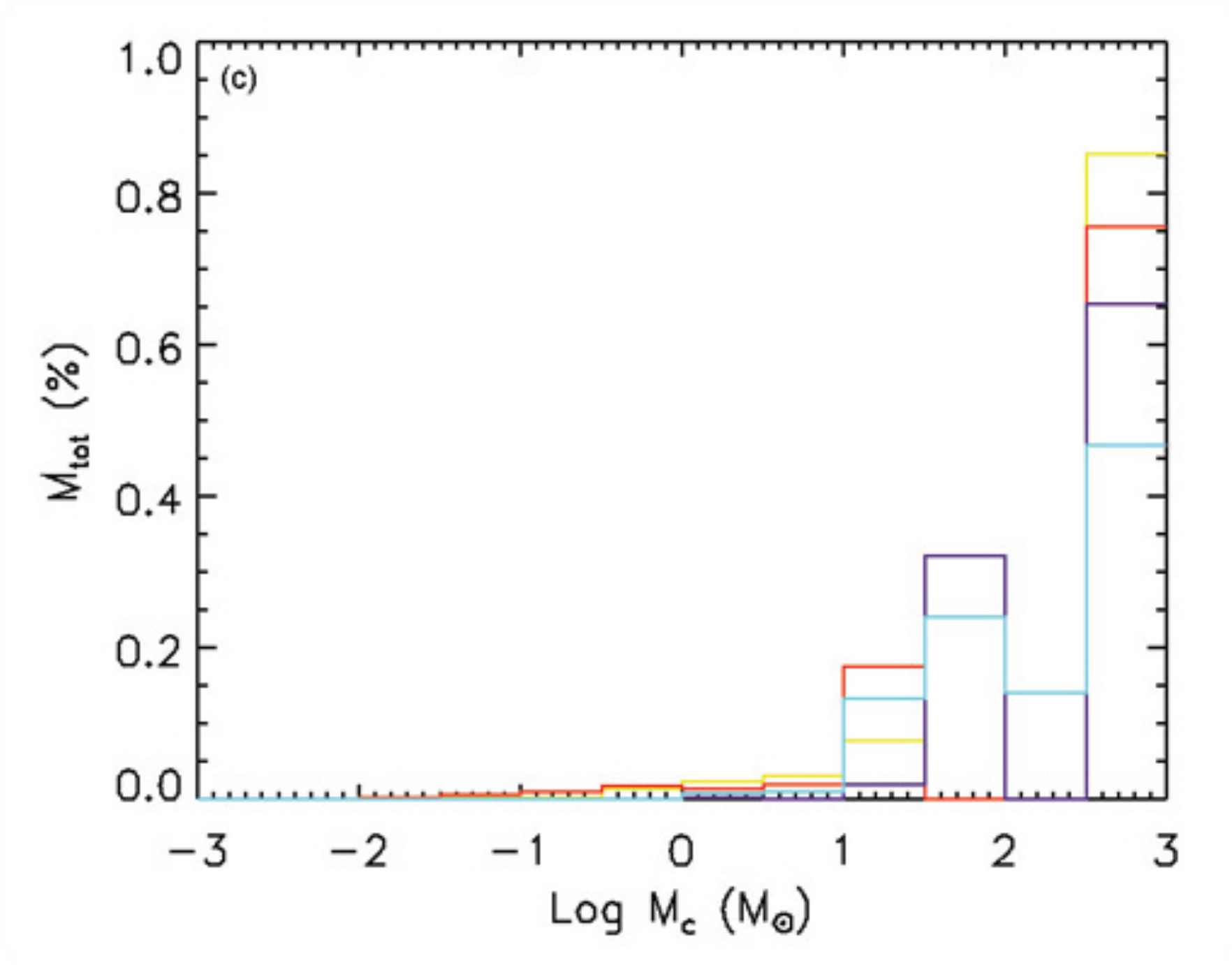}\\
\caption{(a) Velocity histogram of the cloudlets, (b) mass histogram of the
  cloudlets, (c) total cloudlet mass (M$_{\rm tot}$) as function of the cloudlet
  mass. (online: 0.78 pc/cell [navy], 0.39 pc/cell [cyan], 0.20 pc/cell
  [gold], and 0.10 pc/cell [red]).}\label{fig:cloud_prop}
\end{figure}

Figure \ref{fig:cloud_prop} gives velocity (left) and mass (centre)
histograms of the cloudlets , as well as the total mass of the cloudlets as function of the
cloudlet mass (right), at 0.7 Myr for the 4 resolutions shown in Figure
\ref{fig:res_dens}. The massive fragments ($M \gtrsim 10^{2} - 10^{3} ~\rm
M_{\odot}$) present at all resolutions are the remnants of the cloud
core. Since the number of cloudlets increases with resolution, the
highest resolution simulations are comprised of numerous low mass cloudlets. In
general, there is a trend towards increasing smaller mass fragments at all
resolutions. However, in all cases, the majority of mass of the cloudlet
system is found in the massive fragments ($>$ 40~\%) rather than the smaller
fragments. The velocity of the cloudlets does not vary
significantly with resolution. Despite the large number of lower mass
fragments present in the high resolution simulations, they still fall within
the velocity range of $v_{\rm c} = 150 - 400 ~\rm km~s^{-1}$. The bulk of the
cloudlets at all resolutions have a velocity in the range of  $v_{\rm c} = 180 - 
220 ~\rm km~s^{-1}$. This is likely to increase as the evolution progresses and 
the cloudlets are further accelerated by the wind.

Unfortunately, we are unable to fully resolve the interaction of a radiative cloud 
with a supersonic wind at this time. This limits our ability to draw any reliable conclusions 
regarding the small scale evolution of the wind/cloud interaction. A significant increase in 
numerical resolution and/or the use of an adaptive mesh would be required in order for  
convergence to possibly occur. Nevertheless, there is a clear trend in the
large-scale evolution of the wind/cloud 
interaction at all resolutions considered in this study, allowing us to draw some physical 
conclusions. For example, the soft X-ray luminosity is sufficiently resolved and is almost 
constant with numerical resolution (see \S~\ref{xray}). The effect of radiation in keeping 
the cloud cool, suppressing the transverse expansion and minimizing the effect of the 
Kelvin-Helmholtz 
instability, is also not effected by the resolution of the simulation. In all 
cases, the cloud is not immediately destroyed and mixed into the hot wind as seen 
in adiabatic models, but is instead broken-up into numerous small cloudlets. The 
major effect of increasing the resolution is the increased fragmentation of the 
cloud. However, we still see the same qualitative structure at high resolution, 
with the cloud breaking-up to form a filamentary structure that becomes finer and 
finer as more detail is resolved.  

\section{EMISSION IN STARBURST-DRIVEN WINDS}\label{SB_winds}

\subsection{Filamentary H$\alpha$ Emission}

At optical wavelengths (such as H$\alpha$), starburst-driven winds appear as 
spectacular filamentary systems extending several kpc along the minor axis of the 
host galaxy, e.g. M82 \citep{Shopbell1998}, NGC 3079 \citep{Veilleux1994}, NGC 
1569 \citep{Westmoquette2008}. While it has long been proposed that this 
filamentary material was expelled from the central region of the galaxy 
\citep{Lynds1963,Bland1988}, until now the mechanism behind the formation of the 
filaments has not been completely understood. In paper {\sc i}, we proposed that 
the filaments were formed via clouds of disk gas that are broken-up and 
accelerated into the outflow by the ram-pressure of the wind. In order for this 
mechanism to be viable, the cloud fragments need to survive and remain 
sufficiently cool to emit at H$\alpha$ temperatures. The simulations presented in 
this paper allow us to address this important issue.

\begin{figure}[tbp]
\epsscale{0.8}
\plotone{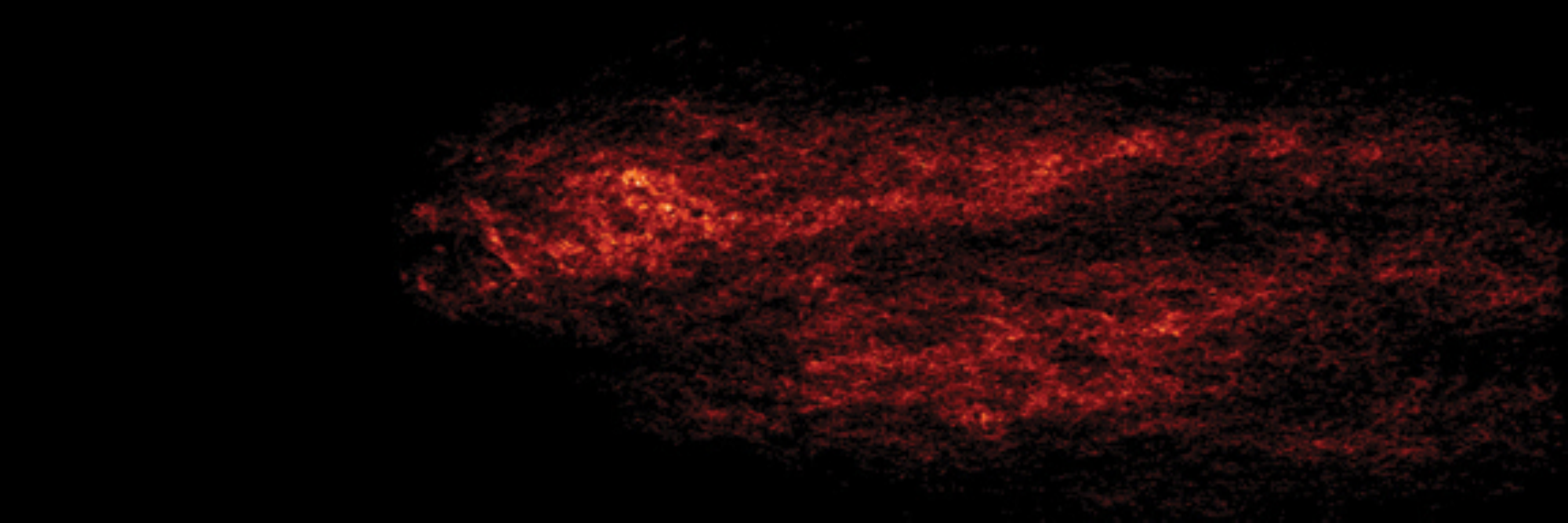}
\caption{Volume rendering of the H$\alpha$ emission in model rf384 at 0.5 Myr 
(Online: movie of the H$\alpha$ emission in model rf384 over a time frame of
  1.37 Myr).}\label{fig:halpha} 
\end{figure}

As in paper {\sc i}, we define the H$\alpha$ emitting gas to be cloud 
material with temperatures in the range of $T = 5 \times 10^3 - 3 \times 10^4
~\rm K$. We note that as photoionization is not included in our model, the 
H$\alpha$ emission discussed in this paper arises solely from shock ionization. 
However, photoionization is known to play a role in the ionization of the 
filaments in many winds. For example, the filaments in M82's wind are known to be 
photoionized at low distances above and below the galactic plane, with shock 
ionization becoming dominant at large distances. An investigation into the effects 
of photoionization on the cloudlets is warranted, but is beyond the scope of
this study. Figure \ref{fig:halpha} shows a three-dimensional volume rendering 
of the density of the H$\alpha$ emitting gas at 0.5 Myr in model rf384 (Online: 
Movie of evolution of the H$\alpha$ emitting gas in rf384 over the first 1.37 
Myr). It can be immediately seen in Figure \ref{fig:halpha} that the H$\alpha$ 
emitting material corresponds to the dense cloud material. Thus, the survival 
mechanism for a cloud proposed in \S~\ref{survival} can be invoked to explain 
the filaments observed in starburst-driven winds.

\begin{figure}[tbp]
\epsscale{1.0}
\plottwo{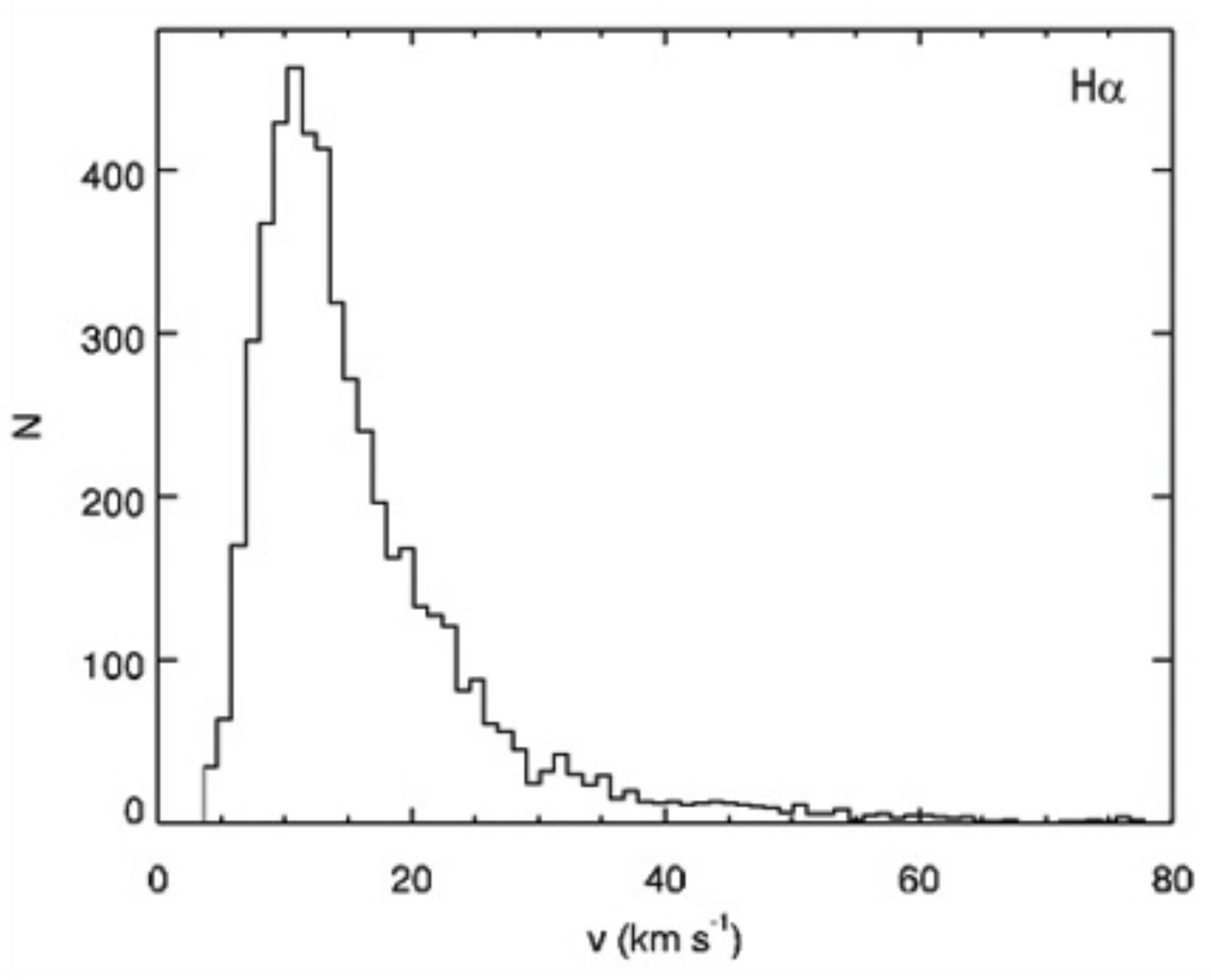}{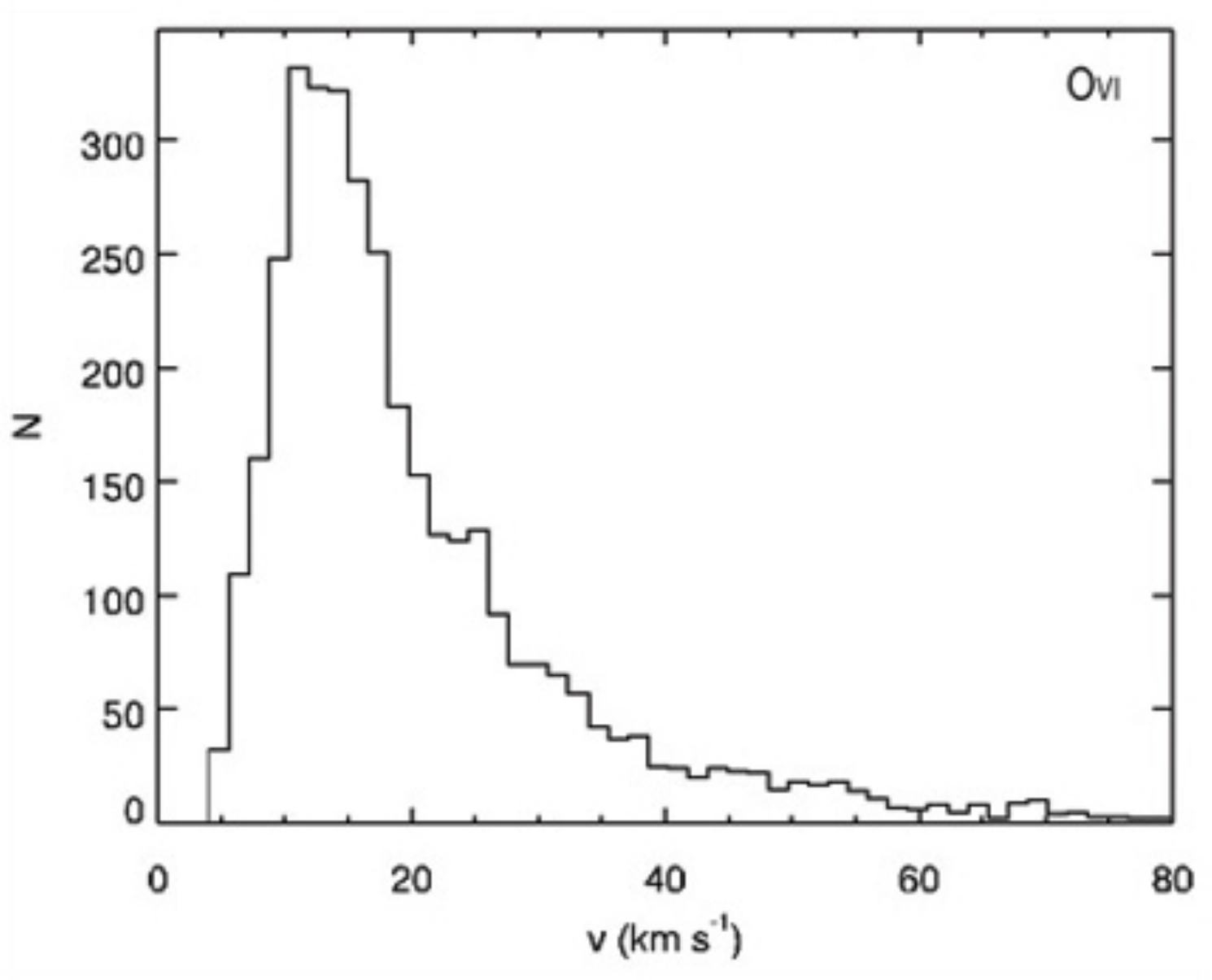}
\caption{Emission weighted histogram of the z-velocity along the filament of the 
H$\alpha$ (left) and O{\sc vi} (right) emission in model rf384 at 0.7 Myr.}\label{fig:vel}
\end{figure}

As discussed in \S~\ref{evolution}, the cloud is broken-up via the 
Kelvin-Helmholtz instability, with the fragments subsequently entrained into the 
outflow forming a filamentary structure. Figure \ref{fig:vel} (left panel) gives 
the emission weighted histogram of the z-velocity along the filament at 0.7 Myr in
model rf384. The majority of the H$\alpha$ emission has velocities in the
range of $v \sim 0 - 30 ~\rm km~s^{-1}$, with the velocity increasing with
distance along the z-axis. This is consistent with the H$\alpha$ gas reaching
higher velocities at larger distances above the galaxy plane, which was observed 
in the global simulations in paper {\sc i}. Note that this result represents 
only the velocity dispersion at the base of a {\em single} filament early in its
evolution, with the velocity likely to increase as the cloudlets, which form the 
filament, are further accelerated in the direction of the flow. 

\begin{figure}[tbp] 
\epsscale{1.0}
\plottwo{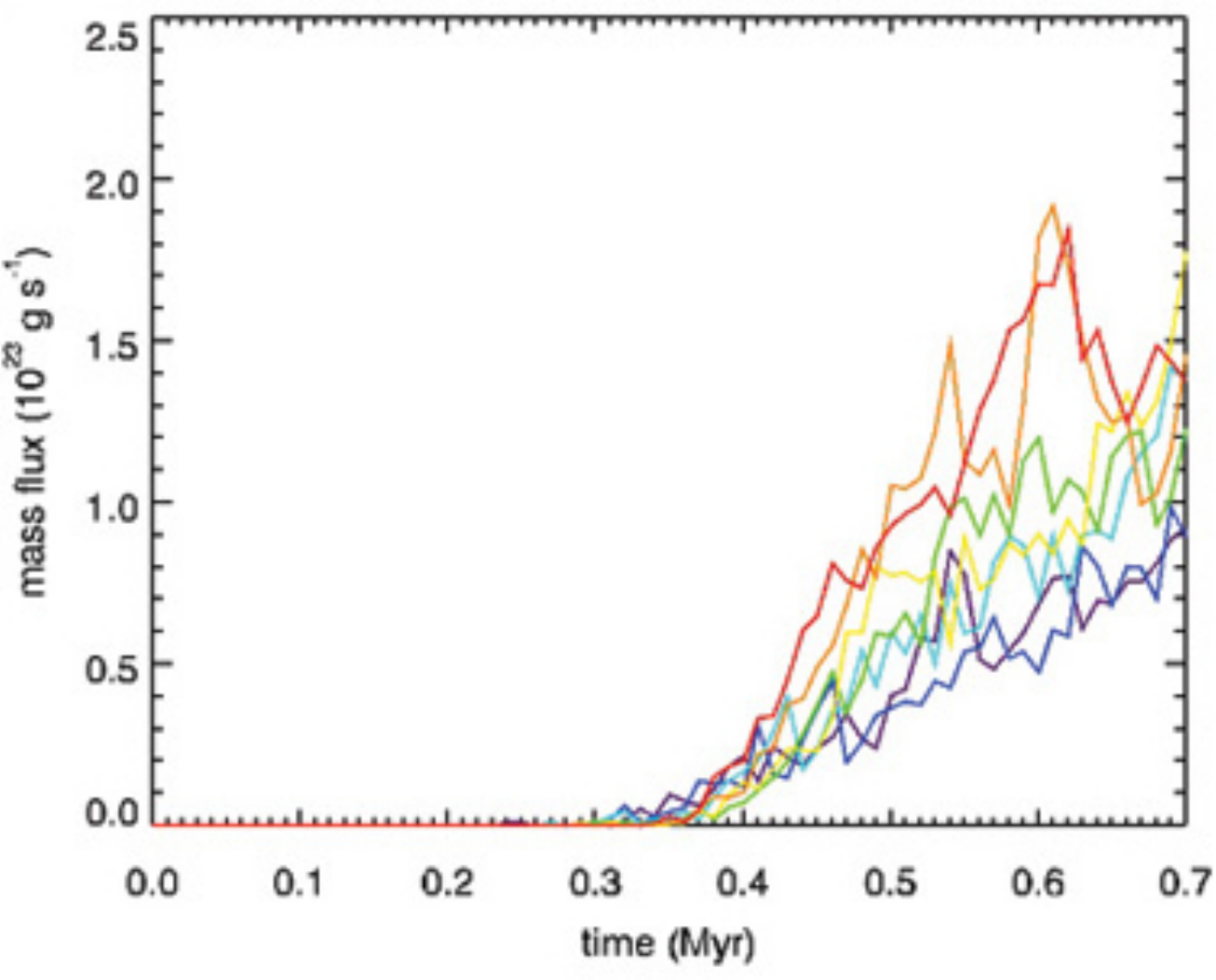}{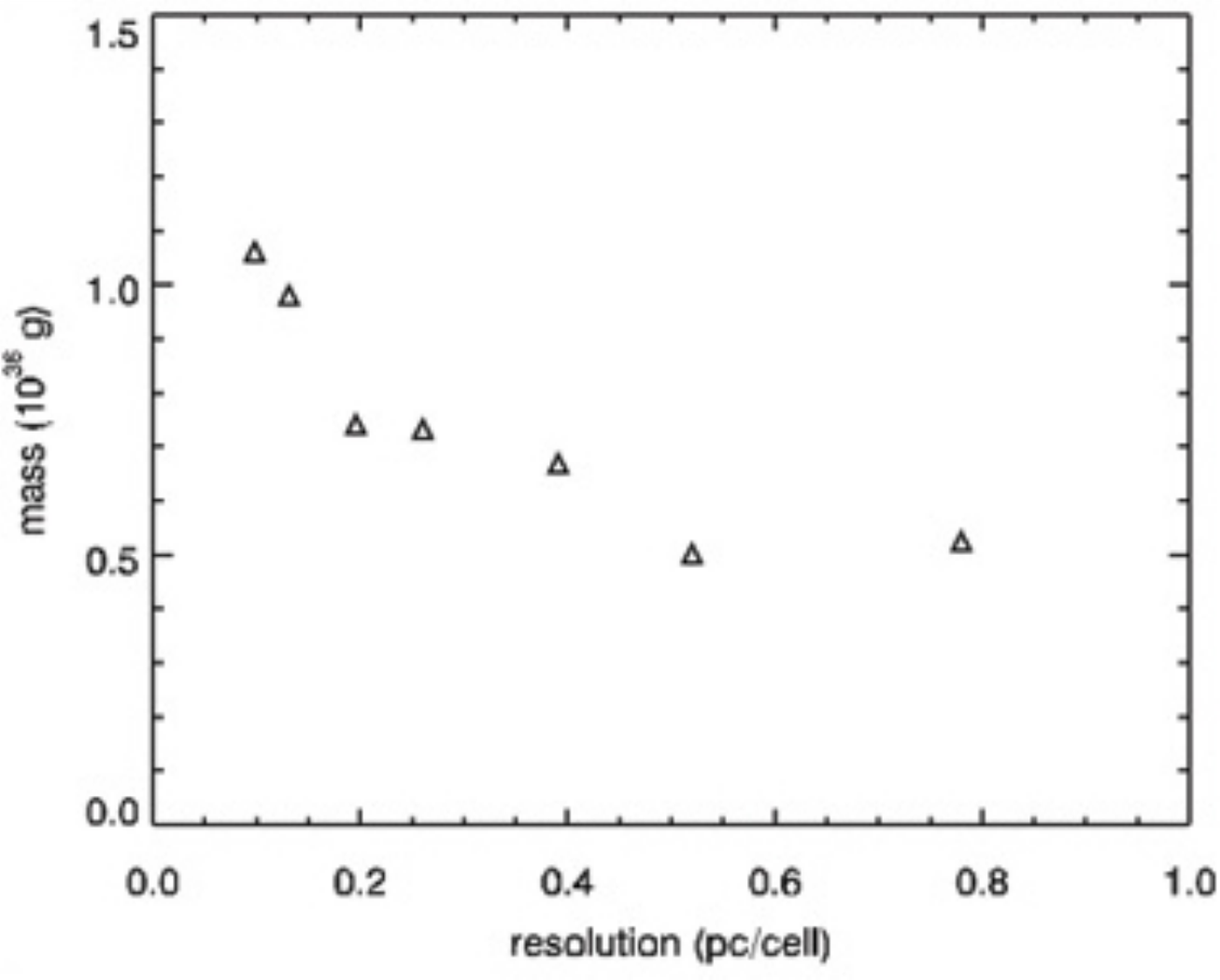}
\caption{Left: mass flux of the H$\alpha$ emitting gas through a surface at 
z=0.75 pc for each resolution (online: 0.78 pc/cell [navy], 0.52 pc/cell [blue], 
0.39 pc/cell [cyan], 0.26 pc/cell [green], 0.20 pc/cell [gold], 0.13 pc/cell 
[orange], and 0.10 pc/cell [red]). Right: total integrated mass flux for the 
H$\alpha$ emitting gas over 0.7 Myr as a function of the numerical 
resolution.}\label{fig:ha_mass_flux}
\end{figure}

As in our global simulations, we again find that it is the ram-pressure of the 
wind that accelerates the clouds. The main difference between the filaments formed 
in paper {\sc i} and the filaments found here is the number of 
cloudlets that comprise the filament; a direct result of the higher numerical 
resolution of the simulations in this work. If we were able to increase the 
resolution of the global simulations, we would find similar fine structure in the 
filaments that are formed. Figure \ref{fig:ha_mass_flux} (left) shows the mass 
flux of the H$\alpha$ emitting gas through a surface at $z = 75 ~\rm pc$ at each 
resolution considered in our study. As expected, there is no H$\alpha$ emitting 
gas passing though the surface until approximately 0.3 Myr, as the dense cloud 
material has yet to encounter the flux surface. As in Figure \ref{fig:mass_flux}, 
the difference in the mass flux at each resolution is a result of the increasing 
fragmentation of the H$\alpha$ emitting clouds at high resolution. The right hand 
panel of Figure \ref{fig:ha_mass_flux} shows the total integrated mass flux, over 
the first 0.7 Myr, of the H$\alpha$ emitting gas as a function of resolution. We 
see an increase in the total mass passing through the flux surface, again caused 
by the increase in fragmentation as the Kelvin-Helmholtz instability is further 
resolved. Even at the high resolution of these simulations the H$\alpha$ mass flux 
is yet to converge.

In \S~\ref{survival} we discussed how a cloud could survive the interaction
with a hot supersonic wind. The ability of a cloud to radiate heat is crucial for 
the clouds survival, allowing it to remain stable to ablation and emit at 
H$\alpha$ temperatures. Without this ability, the cloud quickly heats above $T = 
10^6 ~\rm K$, expands and becomes susceptible to the Kelvin-Helmholtz instability. 
While a radiative cloud is still disrupted, the small sized fragments that are 
broken off the main cloud have cooling times faster than the cloud crushing time 
and the Kelvin-Helmholtz growth rate, and thus, may possibly survive. In the 
adiabatic case, the fragments get heated and destroyed, with the cloud material 
quickly becoming mixed into the hot wind. If the fragments survive, they are 
drawn-out into strands and form a filament downstream of the original cloud 
position, reminiscent of the filaments seen in starburst-driven winds.

\subsection{Soft X-Ray Emission}\label{xray} 

In paper {\sc i} we proposed four mechanisms that could give rise to the soft
X-ray emission that is observed to be spatially correlated to the H$\alpha$
emitting filaments. This correlation has been observed by {\it Chandra} in many 
starburst-driven winds \citep[e.g][]{Cecil2002,Strickland2004a,Strickland2004b}. Here
we summarize the proposed mechanisms:
\begin{enumerate}
\item[(a)] The mass-loaded wind. This is the largest contributor to the soft X-ray 
emission in the global simulations. As mass is ablated from the clouds it is mixed 
into the surrounding turbulent gas, creating a region of hot ($T \gtrsim 10^{6} 
~\rm K$) rapidly cooling gas that emits strongly at soft X-ray energies.
\item[(b)] The intermediate temperature interface between the hot wind and cool 
filaments. Gas at the boundary between the hot and cool gas mixes to produce a 
thin region of intermediate density and temperature. Like the mass-loaded wind, 
this mixed gas is a strong emitter of soft X-rays. 
\item[(c)] Bow shocks. Gas is heated to X-ray temperatures as a bow shock is
formed upstream of dense clouds accelerated into the flow.
\item[(d)] Colliding bow shocks. When two bow shocks interact, the gas is further 
shock heated to X-ray temperatures.
\end{enumerate}
The first two processes involve the mixing of hot and cold gas and could be the 
result of numerical diffusion in the simulations and therefore not physical. Our 
resolution study allows has to examine the effect of increasing resolution on the 
soft X-ray emission and determine the realism of the above possible emission 
processes.

\begin{figure}[tbp]
\epsscale{0.8}
\plotone{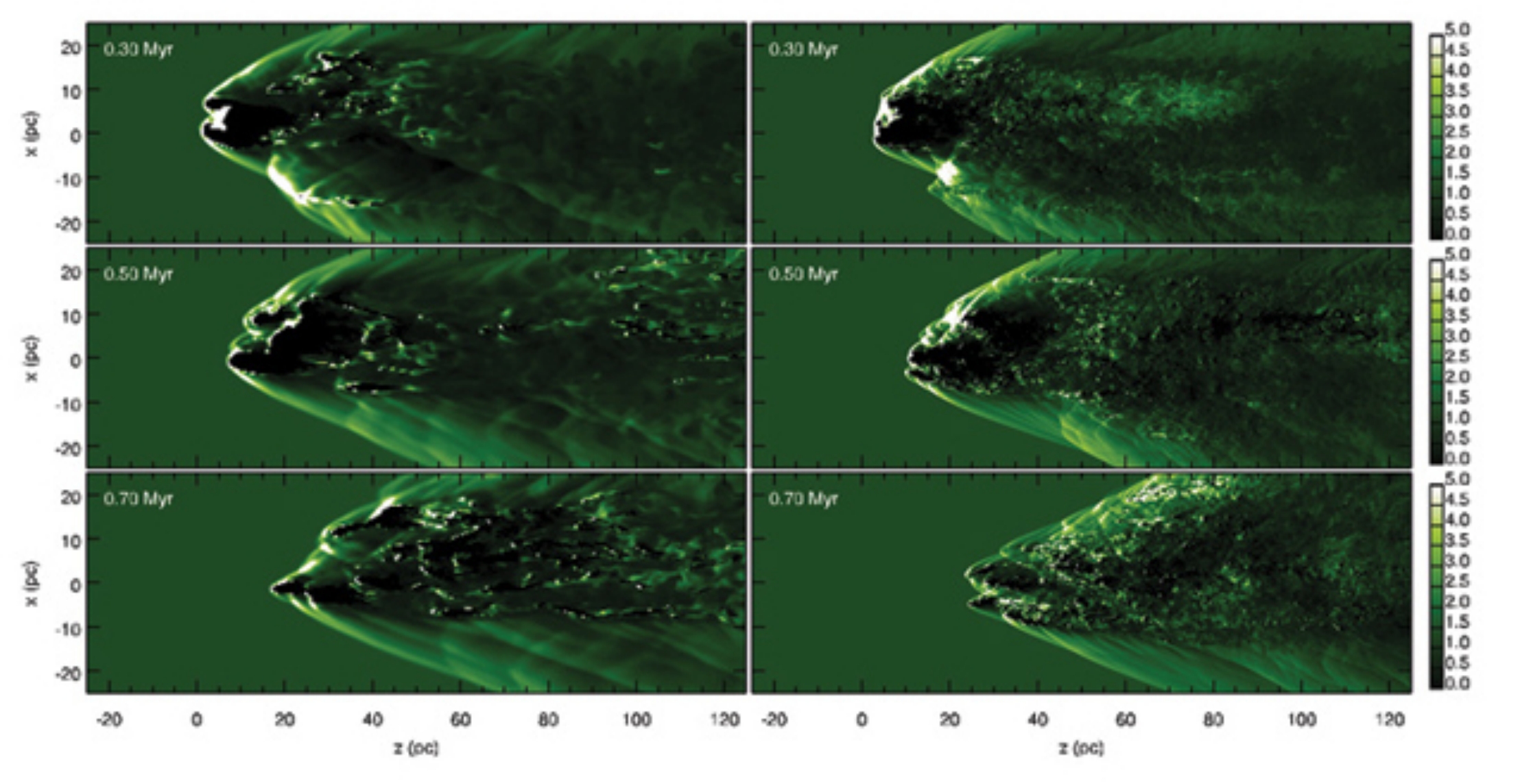}
\caption{Soft X-ray emissivity ($10^{-25} ~\rm erg~s^{-1}~cm^{-3}$) in the y=0 
plane of models rf128 (left) and rf512 (right) at 0.3 (top), 0.5 (middle), and 0.7 
(bottom) Myr epochs.}\label{fig:xray}
\end{figure}

As in paper {\sc i}, we infer the X-ray luminosity in the soft (0.5 - 
2.0 keV) energy band using broadband cooling fractions obtained from the MAPPINGS 
IIIr code \citep{Sutherland1993}. Figure \ref{fig:xray} shows the soft X-ray 
emissivity in models rf128 (left) and rf512 (right) at 0.3, 0.5, and 0.7 Myr 
epochs. The strongest X-ray emitter in both models is the bow shock that 
immediately forms upstream of the cloud as it interacts with the wind. Regions 
where bow shocks are interacting result in the highest X-ray emissivities. Apart 
from a few marginally bright tails coming off some of the cloudlets, particularly 
in model rf128 (see Fig. \ref{fig:xray}; bottom left panels), we see no evidence 
that mass-loading of the wind by ablation from the cloud is a significant 
contributor to the soft X-ray luminosity. We also see very little evidence that 
the intermediate temperature interface plays a significant role. There are a few 
bright regions upstream of cloudlets that have been exposed to the wind. However, 
it is likely that this enhanced emission is related to the X-ray emitting bow 
shocks that have formed around each cloudlet.

The main difference between the simulations shown in Figure \ref{fig:xray} at
high (right panel) and low (left panel) resolution is in the structure of the 
main bow shock and the emission from the cloudlets. While we observe some 
structure in the low resolution simulations, as the resolution is increased we see 
clear regions where colliding bow shocks lead to a significant increase in the 
X-ray emissivity. This is more evident at later times when the main cloud has 
fragmented and there are many X-ray emitting bow shocks upstream of the resulting 
cloudlets (Fig. \ref{fig:xray}; bottom right panels). It might be expected than 
that the increase in fragmentation at higher resolutions would lead to an increase 
in the X-ray luminosity as more bow shocks are formed. However, the majority of 
the cloudlets formed are sheltered from the impacting wind by the main cloud and 
do not form a bow shock. Thus, they are not seen at soft X-ray energies and do not 
contribute to the X-ray luminosity (Fig. \ref{fig:xray}; right panels). 

\begin{figure}[tbp]
\epsscale{0.7}
\plotone{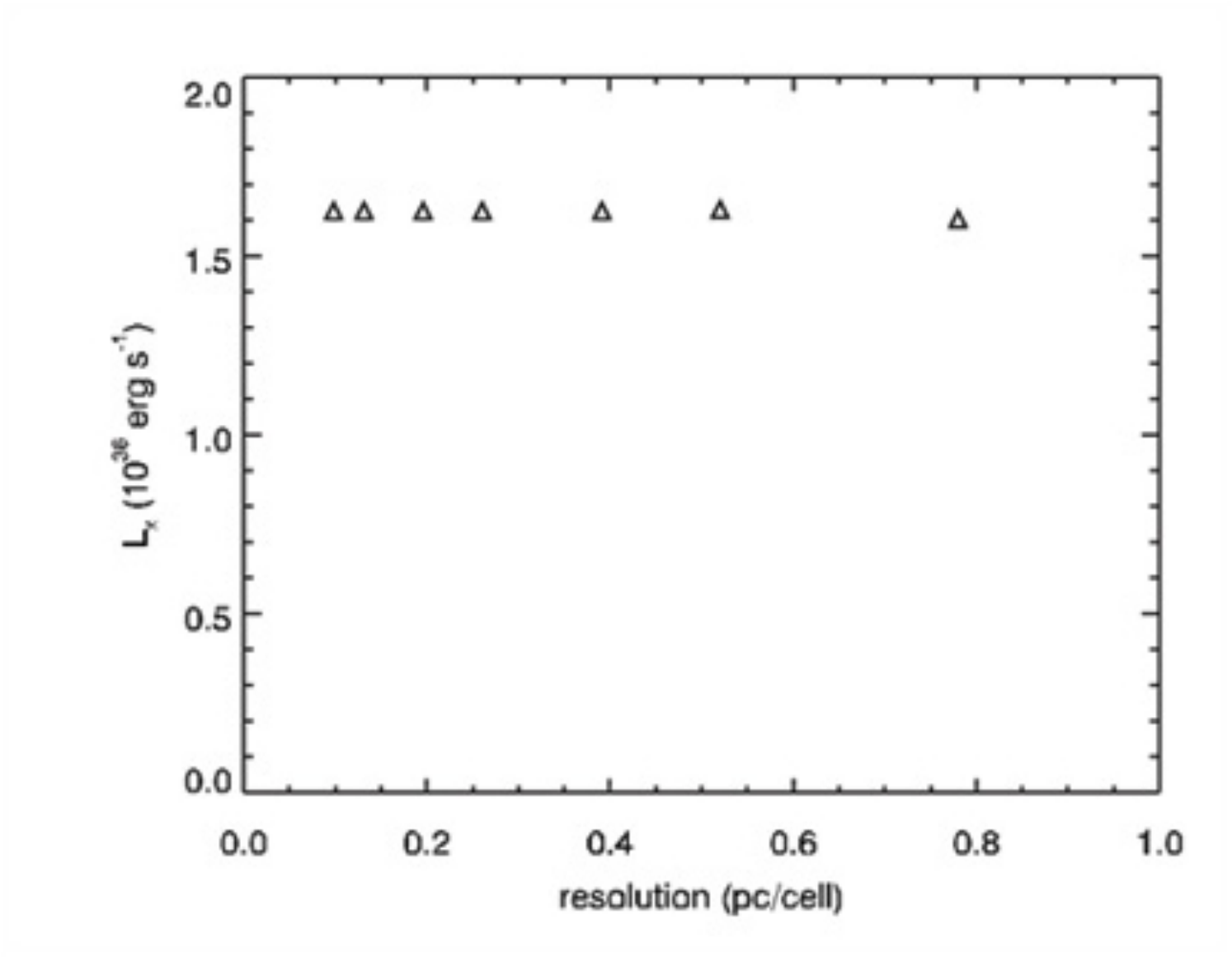}
\caption{Numerical resolution vs total soft X-ray luminosity at 0.7 Myr for the 
radiative fractal cloud.}\label{fig:xray_lum}
\end{figure}

Indeed, the soft X-ray luminosity hardly varies with resolution. Figure
\ref{fig:xray_lum} shows the soft X-ray luminosity as a function of
resolution for our  radiative fractal cloud at 0.7 Myr. At all resolutions, the
luminosity is of  the order of $L_{\rm x} \sim 10^{36} ~\rm erg~s^{-1}$. This 
amount varies only negligibly from low to high resolution. This is a result of the
strongest X-ray source being  the main bow shock upstream of the original gas
cloud, which is  present at all resolutions. As discussed above, we see little
evidence that the mass-loaded component of the wind, nor the intermediate
temperature interface between the hot and cold gas plays a large role in the
soft X-ray emission. While high resolution ($< 0.5 ~\rm pc/cell$) global
simulations are needed in order to confirm this result, we conclude that bow
shocks and their interaction are the main source of the soft X-ray emission in
starburst-driven winds.   

\subsection{O{\sc vi} Emission}\label{Ovi}

\begin{figure}[tbp]
\epsscale{0.5}
\plotone{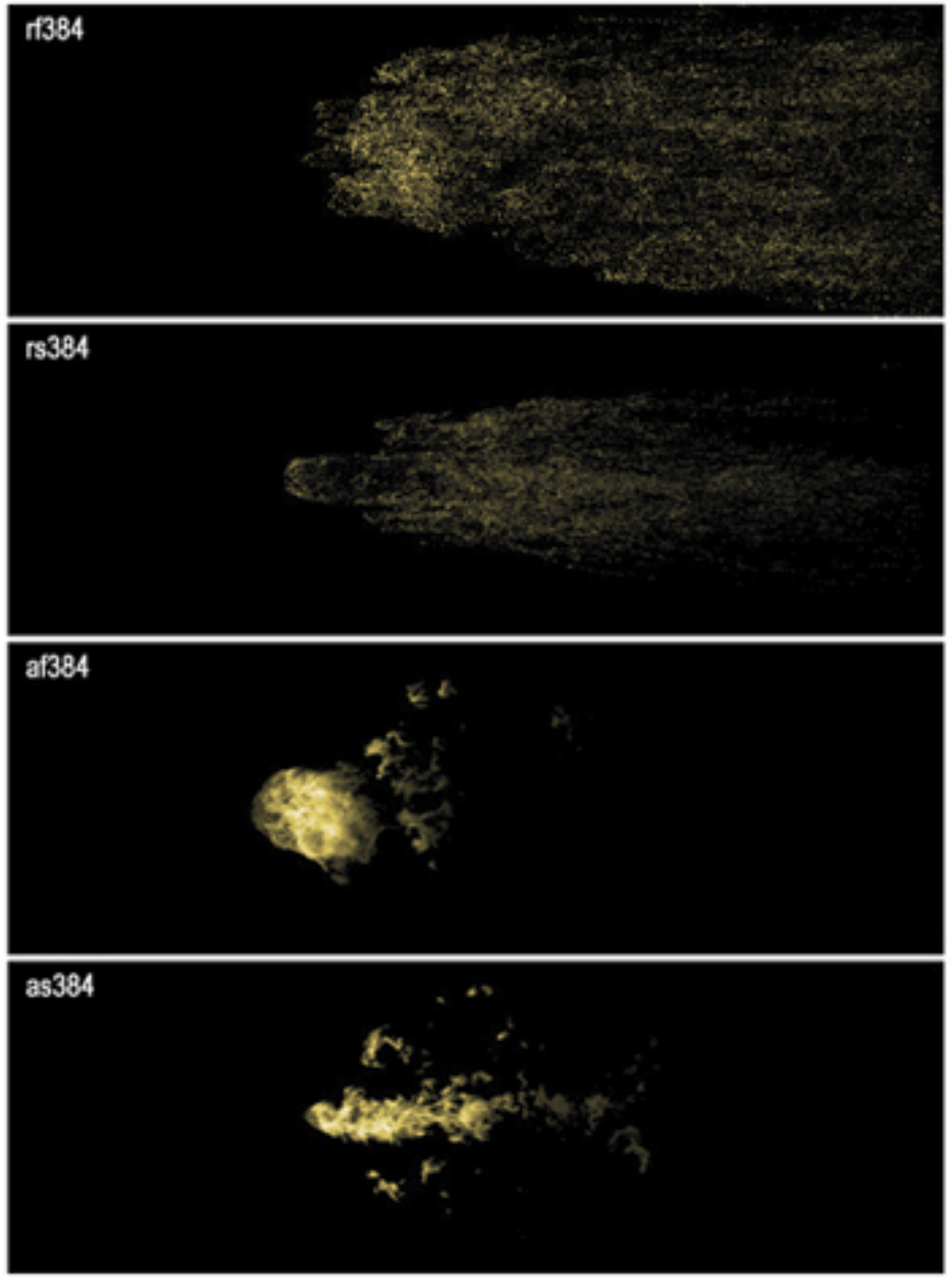}
\caption{Map of the O{\sc vi} emission at 0.7 Myr for both fractal and spherical
  clouds with both radiative cooling included and neglected. Top to bottom:
  model rf384, rs384, af384, and as384.}\label{fig:Ovi}
\end{figure}

The importance of radiative cooling in the formation and survival of the filaments 
was discussed in \S~\ref{survival}. For our proposal of the
formation of a filament, via the break-up and acceleration of cool disk gas
into the wind, to be a viable mechanism, cooling must be present in the
outflow. Observations of the O{\sc vi} emission line could be used to detect this
cooling in the filaments. \citet{Heckman2001} report the detection of O{\sc
  vi} emission in the dwarf starburst galaxy NGC 1705, which they associate
with cooling in the outflow from gas at temperatures of $T \gtrsim 3 \times 10^5
~\rm K$. They propose turbulent mixing layers \citep[e.g.][]{Slavin1993} as a 
possible origin for this emission.

In order to determine were the O{\sc vi} emission may arise in our
simulations, we have produced a ``map'' of the predicted O{\sc vi} emission
in models rf384, rs384, af384, and as384 at 0.7 Myr (Fig. \ref{fig:Ovi}). We 
assume that the O{\sc vi} emission in our simulations falls within the 
temperature range of $T = 1 \times 10^5 - 4 \times 10^5 ~\rm K$. When cooling is 
neglected (bottom two panels), O{\sc vi} emission is only observed around the 
surviving cloud core. This emission is the result of the high degree of mixing of 
the hot and cold gas seen in the adiabatic simulations. In the radiative models 
(top two panels), O{\sc vi} emission is observed throughout the flow and is closely
aligned to the filamentary gas. This emission is caused by the mixing
of hot and cold gas in the vicinity of each cool cloudlet that comprises the
filament. 

The distribution of the O{\sc vi} emission is most significant in the
radiative fractal model (rf384; top panel), where the structure of the
filament can still be clearly seen. The emission weighted histogram of the
z-velocity for this model at 0.7 Myr is shown in the right hand panel of Figure
\ref{fig:vel}. The velocity dispersion of the O{\sc vi} gas is similar to that
of the H$\alpha$ emitting filaments, falling in the range of $v \sim 0 - 40 ~\rm
km~s^{-1}$. While our investigation of the O{\sc vi} emission is only preliminary,
further detection of O{\sc vi} kinematics similar to those proposed in this
work would lend support the premise of cooling in the filaments of starburst 
winds.       

\section{MISSING PHYSICS}\label{missing}

As discussed in \S~\ref{method}, the simulations performed in this study and paper {\sc i} do 
not include thermal conduction or magnetic fields. While an investigation into the role these 
phenomena play in the wind/cloud interaction is beyond the scope of this work, they may 
influence the evolution and survival of the filaments in starburst winds. As such, we briefly 
discuss any differences their inclusion may have had on our results.  

\subsection{Thermal Conduction}

The effects of thermal conduction on the wind/cloud interaction has been
investigated over the last decade by a number of authors
\citep[e.g.][]{Vieser2000,Hensler2002,Marcolini2005,Orlando2005,Recchi2007}. The
immediate effects on the  evolution of a radiative cloud has been described in
detail by \citet{Marcolini2005} who modeled the interaction of a cloud of
radius $R_{\rm c} = 15 ~\rm pc$ and temperature $T_{\rm c} = 10^4 ~\rm K$
with a hot wind of temperature $T_{\rm w} = 10^6 ~\rm K$. Here we summarize
the evolution of a radiative cloud in their two-dimensional simulation that 
includes the effects of thermal conduction:
\begin{enumerate}
\item In the early phases of the cloud's evolution, thermal conduction results
  in a converging shock forming around the cloud. The cloud shrinks in size until
  the central pressure in the  cloud's core is high enough to halt the collapse.
\item The cloud begins to re-expand and over time becomes elongated in the
  downstream direction. 
\item As the cloud evolves, it is compressed in a non-uniform manner due to the shape of the
  cloud and the complexity of the flow.
\item After 1 Myr, the cloud forms a filamentary structure along the symmetry
  axis of the simulation.  
\end{enumerate}
The key difference to the evolution of a cloud in a simulation that neglects thermal 
conduction is that the cloud does not suffer hydrodynamical instabilities and fragment. This 
is because thermal conduction smooths the density and velocity gradients across the surface of 
the cloud, inhibiting the growth of instabilities \citep[also see][]{Vieser2000}. However, 
\citet{Marcolini2005} note that there is also significant mass loss from the cloud through 
evaporation in their thermal conduction models.

\citet{Orlando2005} also considered the effects of thermal conduction on a cloud's
evolution by investigating the survival of a cloud of radius $R_{\rm c} = 1 ~\rm pc$ being 
overrun by Mach 30 and 50 shocks. They performed both three-dimensional simulations that 
ignored radiative cooling and thermal conduction and two-dimensional simulations where they 
were included. Similar to \citet{Marcolini2005}, they found that thermal conduction inhibited 
the growth of hydrodynamical instabilities. In their simulations, the structure of a cloud 
consists of a cool dense core surrounded by a ``corona'' of gas, dominated by thermal 
conduction, that gradually evaporates as the cloud evolves. The cloud does not fragment and 
there is significant mass loss through evaporation. In the case of the slower Mach 30 shock, 
the dense core is unaffected by heat conduction and still fragments into multiple cloudlets. 
They also investigated the X-ray emission that would arise from their simulations 
\citep{Orlando2006}. They found that for the slower shock, the soft X-ray emission arose in 
the thermally conducting corona, while the cold core would likely emit at optical wavelengths. 
In the case of the faster Mach 50 shock, they suggest that there would be no optical component.

In this paper, we find that the soft X-ray emission arises primarily from the main bow shock 
upstream of the original cloud. The optically emitting filamentary gas originates from the 
cool cloudlets that are the remnants of this cloud. As the simulations of 
\citet{Marcolini2005} are the closest in initial conditions to our study, we
use them as a basis for our supposition on the effect of thermal conduction on our results. Most 
significantly, thermal conduction would likely act to inhibit the growth of the 
Kelvin-Helmholtz instability, which is responsible for the high degree of cloud fragmentation 
seen in our simulations. While the initial evolution of the cloud would be similar to that 
discussed in \S~\ref{evolution}, the break-up of the cloud would likely be totally or 
partially suppressed. The initial structure of the cloud (e.g. fractal or spherical) would 
play a role in determining the degree of any fragmentation that occurs. A more cohesive 
H$\alpha$ emitting filamentary structure would be formed. In their study, \citet{Marcolini2005} reported 
the bow shock upstream of the initial cloud to be the main source of the soft X-ray emission 
in their non thermal conducting simulations. When thermal conduction was considered they found 
an increase in the soft X-ray emission in the region between the cloud surface and the bow 
shock. It is likely that our simulations would show a similar result. Note that this emission 
is still complimentary to the filamentary H$\alpha$ emission, as seen by observations of 
starburst winds.

We conclude that thermal conduction may help the survival time of a radiative cloud overrun by 
a supersonic wind by stabilizing it against the destructive effect of the Kelvin-Helmholtz 
instability. While there will be less mixing of cloud material into the wind by hydrodynamical 
effects, there will still be significant mass loss from the cloud via evaporation. However, as 
discussed below, the presence of a magnetic field may counteract any benefits to cloud 
stability gained through thermal conduction.    

\subsection{Magnetic Fields}

Magnetic fields have been shown by a number of authors to have a varying effect on the 
wind/cloud interaction. A variety of different circumstances have been taken into 
consideration, such as the strength and orientation of the magnetic field. Early models were 
largely two-dimensional and ignored the effects of radiative cooling 
\citep[e.g.][]{MacLow1994}, while later models have been three-dimensional 
\citep[e.g.][]{Gregori1999,Gregori2000,Shin2008}, and/or included radiative cooling and 
thermal conduction \citep[e.g.][]{Fragile2005,Orlando2008}. These studies have shown that the 
orientation of the magnetic field (i.e. parallel, perpendicular, or oblique to the flow) has a 
significant effect on the evolution of the cloud. For example, \citet{Shin2008} found that in 
the case of a strong oblique magnetic field, the cloud was actually pushed out of the central 
y=0 plane as it evolved. It has also been suggested that magnetic fields can help suppress the 
hydrodynamical instabilities that shred the cloud in a non-magnetized simulation 
\citep{MacLow1994,Fragile2005}, with less fragmentation occurring with increasing field 
strength. On the other hand, \citet{Gregori1999} show that in a three-dimensional model, the 
growth of hydrodynamical instabilities can actually be accelerated by the presence of a 
magnetic field.

It is widely thought that the presence of a magnetic field will act to suppress thermal 
conduction \citep[e.g.][]{Chandran1998,Narayan2001}. \citet{Marcolini2005} suggest that the 
coefficient of thermal conductivity $\kappa$ may be significantly reduced below the Spitzer 
level in the presence of a magnetic field. To investigate how a reduced degree of conductivity 
may effect their results, they performed a number of simulations where $\kappa$ was below the 
standard Spitzer level $\kappa_{\rm sp} = 6.1 \times 10^{-7} ~T^{5/2} ~\rm erg~s~K^{-1}$. They 
found that at sub-Spitzer levels, thermal conduction is not as efficient at suppressing the 
hydrodynamical instabilities that fragment the cloud. More recently, \citet{Orlando2008} ran a 
series of two-dimensional axisymmetric simulations that included radiative cooling, thermal 
conduction, and magnetic fields. They showed that while thermal conduction was not completely 
suppressed in the presence of a magnetic field, regardless of the orientation of the 
field, the cloud would be broken up and destroyed by hydrodynamical instabilities. However, 
they also found that this effect was lessened in the case of a strong magnetic field. 

While clearly the interplay between thermal conduction and magnetic fields will influence the 
evolution and survival of the filaments in starburst winds, without detailed three-dimensional 
simulations (with initial conditions appropriate to the problem) that consider radiative 
cooling, thermal conduction and the complex nature of magnetic fields, it is difficult to 
hypothesize what effect the magnetic field would have on a filament's formation and survival. 
Nevertheless, a filamentary structure is still likely to be formed by the wind/cloud 
interaction, with the most significant difference being in the number of cloudlets that 
comprise the filament. We do not anticipate that there would be a large effect to the 
H$\alpha$, soft X-ray, and O{\sc vi} emission processes discussed in \S~\ref{SB_winds}.     

\section{SUMMARY}\label{summary}

We have performed a series of three-dimensional simulations of the interaction
of a supersonic wind with a radiative cloud. We consider two different
cloud geometries (i.e. fractal and spherical), which enable us to
investigate the impact of the initial shape and structure of the cloud
on its subsequent evolution. This work was motivated by the 
simulations of the formation of a starburst-driven galactic wind in a
inhomogeneous interstellar medium, reported in paper {\sc i}. The 
aim of this work is to investigate the possible survival mechanism of a cloud 
accelerated by a hot freely expanding wind. We also set out to determine the 
effect of the numerical resolution of the evolution of the cloud and the implied 
soft X-ray emission associated with the interaction. The results of this study are 
as follows:
\begin{enumerate}
\item Both the initial geometry and the density distribution of the cloud
  significantly affect its evolution. A cloud which has a more inhomogeneous
  distribution of density fragments more than a cloud with a more
  uniform structure (e.g. a sphere). The wind more rapidly breaks the cloud apart 
  in regions where it encounters the least density.

\item A radiative cloud survives longer than an identical adiabatic
  cloud. This is a result of the lower degree of heating in the radiative
  cloud, which suppresses the transverse expansion seen in the adiabatic
  case. The radiative cloud experiences a lower degree of acceleration and
  has a higher relative Mach number to the flow, diminishing the destructive
  effect of the Kelvin-Helmholtz instability.

\item The number of fragments formed by the break-up of the cloud increases as
  a power law with increasing numerical resolution. This is a direct result
  of further resolving the Kelvin-Helmholtz instability, which grows more quickly 
  at shorter wavelengths. The number of fragments formed increases
  down to the resolution of the simulation and will not converge.

\item The calculated mass flux increases with numerical resolution. This is
  due to the turbulent nature of the stream and the increasing fragmentation
  of the cloud. High ($ < 0.1~ \rm pc/cell$) resolution and an adaptive mesh would
  be required for convergence to possibly occur.

\item A radiative cloud fragments into numerous cool, small dense
  cloudlets. These cloudlets are entrained into the turbulent flow,
  forming an overall filamentary structure, with regions where the concentration of
  cloudlets is higher. The velocity of the cloudlets at 0.7 Myr falls in the range 
  of $v_{\rm c} = 150 - 400 ~\rm km~s^{-1}$ irrespective of resolution. 

\item The filamentary structure that is formed and the range of velocities found
  are in good agreement with optical observations of starburst-driven
  winds. Thus, we confirm our conclusion from paper {\sc i}, that H$\alpha$ 
  emitting filaments can be formed from clouds accelerated into a supersonic wind 
  by the ram-pressure of the wind.
  
\item There is little variation in the estimated soft X-ray luminosity of the 
radiative fractal cloud at all numerical resolutions considered, indicating that 
the X-ray emission is well resolved.

\item Soft X-ray emission arises primarily from the main bow shock, produced in
  the initial interaction, and from bow shocks produced upstream
  of fragments that are directly exposed to the wind. Regions where these bow shock
  interact are strong X-ray emitters. We see little evidence that the mixing
  of hot and cold gas (e.g. mass-loading and the boundary between the cool
  cloud material and the hot wind), contribute significantly to the X-ray
  emission.
  
\item The O{\sc vi} emission arises is the same vicinity as the H$\alpha$ emission 
and has comparable emission weighted velocities, suggesting that the detection 
of O{\sc vi } in an outflow may be indicative of cooling in the filaments. 

\end{enumerate} 

The ability for a cloud to radiate heat is crucial for it to survive immersed 
inside a hot, turbulent, supersonic wind. While effects such as thermal conduction 
and magnetic fields will have an effect on the clouds survival, without radiative 
cooling the cloud is quickly destroyed by the Kelvin-Helmholtz instability, with 
the cloud's material completely mixed into the surrounding stream. Thus, for a 
model of the wind/cloud interaction to be realistic, radiative cooling certainly 
cannot be neglected. 

Through the results of this work and paper {\sc i}, we have shown that an
optically emitting filament is easily formed via the interaction of a cool,
dense cloud and a hot, tenuous supersonic wind under the conditions typically
found in a starburst wind. We also find soft X-ray emission that has a natural
spatial relationship to the filamentary gas. A relationship also seen in
Chandra observations of these winds. Clearly, the multiphase nature of the
interstellar medium is crucial for the formation of the filaments in starburst
winds and can help explain much of the optical and soft X-ray emission
detected in these complex objects.     

\bibliographystyle{apj}

\end{document}